\documentclass[%
reprint,
superscriptaddress,
amsmath,
amssymb,
aps,
]{revtex4-2}
\usepackage{graphicx}
\usepackage{dcolumn}
\usepackage{bm}
\usepackage[hidelinks]{hyperref} 
\hypersetup{
	colorlinks = true,
	linkcolor = blue,
	anchorcolor = blue,
	citecolor = blue,
	urlcolor = blue
}
\usepackage{enumitem} 
\usepackage{makecell}
\usepackage{newtxtext,newtxmath}
\usepackage{enumitem} 
\usepackage{makecell}
\usepackage{lineno}
\usepackage[caption=false]{subfig}
\usepackage{multirow}
\usepackage{booktabs}
\usepackage{threeparttable}
\usepackage{cuted}
\newcommand{\aap}{Astronomy \& Astrophysics}
\newcommand{\apjl}{Astrophysical Journal Letters}
\newcommand{\mnras}{Monthly Notices of the Royal Astronomical Society}
\newcommand{\aj}{Astronomical Journal}
\newcommand{\actaa}{Acta Astronomica}
\newcommand{\apss}{Astrophysics and Space Science}
\usepackage[T1]{fontenc}

\begin{document}
	\preprint{APS/123-QED}	
	\title[The evolution of new-born magnetars]{Spin, inclination, and magnetic field evolution of magnetar population in Vacuum and plasma-filled Magnetospheres}
	
	\author{Jun-Xiang Huang}
	\affiliation{Guangxi Key Laboratory for Relativistic Astrophysics, School of Physical Science and Technology, Guangxi University, Nanning 530004, China}
	
	\author{Hou-Jun L\"{u}}
	\email{lhj@gxu.edu.cn}
	\affiliation{Guangxi Key Laboratory for Relativistic Astrophysics, School of Physical Science and Technology, Guangxi University, Nanning 530004, China}

        \author{Jared Rice}
	\affiliation{Department of Mathematics and Physical Science, Southwestern Adventist University, Keene, Texas 76059, USA}
	
	\author{En-Wei Liang}
	\affiliation{Guangxi Key Laboratory for Relativistic Astrophysics, School of Physical Science and Technology, Guangxi University, Nanning 530004, China}
	
	\date{\today}

\begin{abstract}
	Magnetars are potential energy sources or central engines for numerous transient phenomena in the Universe. How newborn magnetars evolve in different environments remains an open question. The majority of previous studies on the evolution of  newborn magnetars considered either the spin evolution, inclination evolution, or magnetic field evolution in vacuum or in a plasma-filled magnetosphere. Based on both observed and candidate magnetars, it is found that the periods of all magnetars or candidates appear as a bimodal distribution, and are defined as the “long-P” and “short-P” magnetar subclasses, respectively. We find that for the “short-P” subclass of magnetars, the $\dot{P}$ values also appear as a bimodal distribution, and therefore can be classified as “high-$\dot{P}$ short-P” and “low-$\dot{P}$ short-P” magnetar subclasses. In this paper, we use Monte Carlo simulations to generate synthetic magnetar populations and investigate the evolution of the “high-$\dot{P}$ short-P” and “low-$\dot{P}$ short-P” magnetar subclasses by considering both the magnetar spin and inclination, as well as the decay of their magnetic field within their evolution in both vacuum and plasma-filled magnetospheres. We find that the magnetar evolution is dependent on both spin and magnetic field, but seems to be insensitive to inclination evolution and magnetospheric environment for the “high-$\dot{P}$ short-P” subclass. In comparison for the case of “high-$\dot{P}$ short-P”, the magnetar evolution is dependent on spin, magnetic field, and inclination evolution, as well as the magnetospheric environment. The best evolution model should be the case of inclination evolution in vacuum with a small value of $\overline{\mathrm{FOM}}$. The differences in the best-fit parameters also suggest that the “high-$\dot{P}$ short-P” and “low-$\dot{P}$ short-P” magnetar subclasses may be tracking with different evolution channels.
\end{abstract}

\maketitle

\section{Introduction}\label{sec:intro}
	 Magnetars are young, strongly magnetized neutron stars and have been studied extensively in the last four decades\citep{Daugherty.Harding.1983, Usov.1984, Paczynski.1992, Duncan.et.al.1992}. Magnetars as central engines have been utilized to explain some observational phenomena such as soft gamma repeaters (SGRs) \citep{Kaspi.et.al.2017}, gamma-ray bursts (GRBs) \citep{Zhang.2018}, fast radio bursts (FRBs) \citep{Zhang.2023}, x-ray transients \citep{Xue.et.al.2019}, as well as anomalous x-ray pulsars (AXPs) \citep{Paradijs.et.al.1995}. In recent years, a wide range of transient phenomena such as short bursts, outbursts, and giant flares have been suggested to originate in magnetar central engine activity \citep{Ersin.et.al.1999, Ersin.et.al.2000,Woods.et.al.2004,Hurley.et.al.2005}. Magnetars exhibit interesting temporal behaviors such as enhanced spin-down, glitches, and antiglitches \citep{Dib.et.al.2014, Woods.et.al.1999, Sinem.et.al.2014}. So far more than 30 magnetars have been discovered and all essential information about them is available in the online McGill magnetar catalogue \footnote{\url{http://www.physics.mcgill.ca/~pulsar/magnetar/main.html}} \citep{Olausen.et.al.2014}. More detailed summaries of the observed properties of magnetars can be found in the literature reviews \citep{Woods.et.al.2006, Kaspi.2007, Mereghetti.2008, Hurley.2011b, Rea.et.al.2011}.
	
	From the observational point of view, the spatial distribution of magnetars in our Galaxy is close to the Galactic plane with a scale height of $20-30$ parsecs (pc), and the measured average magnetar spatial velocity of 200 $\mathrm{km~s}^{-1}$ indicates that the magnetar lifetime upper limit can be estimated as 100 kyr \citep{Tendulkar.et.al.2013}. On the other hand, magnetars that are associated with supernova remnants suggest that the lifetime is younger than that of normal neutron stars (\cite{Ferrario.et.al.2006}). However, the observed x-ray magnetar pulsation periods are longer than 2 s, which suggests that the magnetar spin periods are obviously longer than those of other neutron stars (see Table.~\ref{tab:observation}). The large period and younger lifetime of magnetars place them into a particular region on the $P-\dot{P}$ diagram \citep{Manchester.et.al.2005}. The rapid loss of magnetar rotational energy during their short lifetime indicates  the possibility of the existence of another braking mechanism such as magnetic braking \citep{Colpi.et.al.2000,Vigan.et.al.2013}. Within this scenario, the high surface dipole field strength ($\sim10^{14}-10^{15}$ G) can slow down the NS rotational period from milliseconds to seconds during the typical lifetime of a supernova remnant. Such a strong magnetic field is thought to be a fossil field remnant of the progenitor star via magnetic flux conservation. It can be amplified by a dynamo mechanism such as the Tayler-Spruit dynamo, which is driven by the Tayler instability \citep{Spruit.2002}. The dynamo action with efficient convective mixing can also amplify the magnetic field, but it requires a rapid rotation with a period as fast as $\sim$1 ms \citep{Usov.1992,Duncan.et.al.1992,Thompson.et.al.1993,Thompson.et.al.1995,Thompson.et.al.1996}. Such millisecond newborn magnetars are suggested to be GRB central engines (\citealt{Usov.1992}; \citealt{Troja.et.al.2017,Lv.&.Zhang.2014,Lv.2015}), but there remains a lack of direct observational evidence of this \citep{Zhang.2018}. 
	
	The initial spin periods and magnetic fields as well as the magnetar inclination angles are essential parameters necessary in understanding magnetar physical processes. Moreover, the evolution of those three parameters may provide  details of the magnetar distribution and evolution. For example, \cite{Gill.et.al.2007} adopts the magnetic field decay to explain the clustering of AXPs in the $P-\dot{P}$ diagram. Reference \cite{Jawore.et.al.2020} further simulated the distribution of magnetar populations in the $P-\dot{P}$ diagram by accounting for magnetar``fade-away" when the magnetic field is not strong enough to power the pulsed emission. However, previous studies on pulsar spin evolution focus on ideal conditions with a vacuum magnetosphere \citep{Deutsch.et.al.1955,Goldreich.et.al.1970}, which is inconsistent with observations of the braking index of pulsars observed in our Galaxy \citep{Archibald.et.al.2016,Lower.et.al.2021} and the braking index of nascent NSs in GRBs \citep{Lasky.et.al.2017,lv.et.al.2019,Xiao.et.al.2019}. 
 
   Reference \cite{Goldreich.et.al.1969} suggested that a rotating neutron star may be surrounded by plasma instead of vacuum. If this is the case, the charged particles in the plasma would be extracted along the magnetic field and naturally populate the magnetar's exterior, and the charged plasma would alter the magnetospheric structure and affect the magnetar spin evolution \citep{Harding.1999}. Several groups have calculated the shape of the magnetosphere with arbitrary inclination, and determined that the spin-down power of a millisecond pulsar in plasma differs from that of in vacuum \citep{Spitkovsky.et.al.2006, Timokhin.2006, Kalapotharakos.Contopoulos.2009}. On the other hand, observations of radio pulsars also support the differences  discussed above \citep{Kramer.et.al.2006, Lorimer.et.al.2012, Camilo.et.al.2012}. For example, \cite{Gullon.et.al.2014} simulated a population synthesis of isolated radio pulsars by adopting magnetospheric models with magnetic field and inclination evolution, and found that magnetars with a spin period of less than 0.5 s are consistent with the observational data.

   The majority of previous studies employed a constant inclination in their simulations of magnetar evolution, even though the inclination angle is related to the magnetar spin evolution in different magnetospheric environments \citep{Philippov.et.al.2014}. Within plasma-filled magnetospheres, the coupling between inclination angle and spin-down evolution is different from that of in vacuum. The spin-down of the magnetar would cease entirely at alignment rotation ($\chi=0^{\circ}$) in vacuum. However, due to the complexity of solving the magnetospheric structure analytically or in the absence of self-consistent models, the time evolution of pulsar inclination is usually ignored \citep{Faucher.et.al.2006}.
	
	Understanding the properties of newborn magnetars remains an open question because of the uncertainty of the evolutionary environment and the lack of direct observational evidence of newborn magnetars. The majority of previous studies on the evolution of newborn magnetars consider either spin or inclination evolution, or magnetic field evolution in vacuum or in plasma-filled magnetospheres. In this paper, by adopting Monte Carlo simulations to generate synthetic magnetar populations, we investigate the evolution of newborn magnetars by considering all spin and inclination angle values as well as the decay evolution of the magnetar magnetic field in both vacuum and plasma-filled magnetospheres. Moreover, we also try to reproduce the distribution of magnetars in the $P-\dot{P}$ diagram, and then compare with observed magnetars. In Sec. 2, we briefly introduce the coupling between magnetar spin and inclination evolution in both vacuum and plasma-filled magnetospheres. The sample selection and related statistical analysis of observed magnetars or candidates are presented in Sec. 3. The details of the simulations are shown in Sec. 4. In Sec. 5, we simulate the magnetar spin evolution and find the best-fit parameters for the fixed and coupled inclination evolution models in both vacuum and plasma-filled magnetospheres, respectively, and then compare with observations. The conclusions are drawn in Sec. 6 with further discussion.

\section{Spin and inclination evolution of magnetars}
	\subsection{Vacuum magnetosphere}
	A spheroidal neutron star will brake via an external electromagnetic torque component opposite to the spin direction ($\bf{\Omega}$), which can produce a spin evolution obeying the following equation \citep{Philippov.et.al.2014}:
        \begin{equation}
		\label{dOmega}
		\dot{\Omega}_\mathrm{vac}(t)=-\frac{2 R^{6}}{3 I c^{3}} B(t)^{2} \sin ^{2} \chi(t) \Omega^{3}(t),
	\end{equation}
	where $R$, $I$, and $c$ are the radius and moment of inertia of the spherical star, and the speed of light, respectively. $B$ is the surface magnetic field at the dipole cap, $\chi$ is the inclination of the magnetic axis $\bf{\mu}$ relative to the rotation axis $\bf{\Omega}$, and $\Omega$ is the angular frequency. The component of the external torque that is perpendicular to the spin direction $\bf{\Omega}$ will cause deflection of the magnetic axis $\bf{\mu}$. Thus, the inclination angle evolution can be expressed as \citep{Philippov.et.al.2014}
        \begin{equation}
		\label{dchi}
		\dot{\chi}_\mathrm{vac}(t)=-\frac{2 R^{6}}{3 I c^{3}} B(t)^{2} \sin \chi(t) \cos \chi(t) \Omega^{2}(t).
	\end{equation}
	  
	Given the magnetar evolution time scale, which is longer than thousands of years, the decay of the magnetic field due to diffusion and Ohmic dissipation cannot be ignored \citep{Geppert.et.al.1994,Tauris.et.al.2001}. Here, we adopt a power-law decay form to simulate the evolution of field strength with time \citep{Beniamini.et.al.2019}
        \begin{equation}
		\label{Bt}
            B(t)= B_{0} (1+\frac{\beta t}{\tau_{B}})^{-1/\beta},
        \end{equation}
	where $B_0$ and $\tau_{B}$ are the initial surface magnetic field and the characteristic decay time scale of the magnetic field, respectively. The $\beta$ is a free parameter, and the form of the magnetic field decay is close to exponential $B(t)=B_{0} e^{-t/\tau_{B}}$ when it is near zero.
    
    Together with Eqs.~(\ref{dOmega}) and (\ref{dchi}), one has
        \begin{equation}
		\label{Omega_cos_chi}
		\Omega\cos{\chi}=\Omega_0\cos{\chi_0},
	\end{equation}
	where $\Omega_0$ and $\chi_0$ are the initial angular frequency and inclination angle, respectively. The spin evolution, which can be found from the derivative of the period in Eq. (\ref{dOmega}), can be rewritten as
        \begin{equation}
		\label{dP}
		\dot{P}_\mathrm{vac}(t)=\left(\frac{P_{0}^{2}}{P(t)^{2} \cos ^{2} \chi_{0}}-1\right) \frac{P(t) \exp ({-2 t/\tau_{B})}}{\tau_{\chi}},
	\end{equation}
	where 
        \begin{equation}
		\tau_{\chi}=\frac{3 c^{3} I P_{0}^{2}}{8 \pi^{2} R^{6} B_{0}^{2}\cos ^{2}\chi_{0}}
	\end{equation}
	is the alignment time scale of a vacuum pulsar, and $P_{0}$ is the initial spin period.
	
	By solving Eq.~(\ref{dP}), one can derive the analytic solution of the spin period evolution,
        \begin{equation}
		\label{Pt}
            \scalebox{0.95}{$P_\mathrm{vac}(t)=\left\{\begin{array}{ll}
            P_{0} \sqrt{1+\left\lbrace 1-\exp \left[\frac{\tau_{B}}{\tau_{\chi}}\left(e^{-\frac{2 t} {\tau_{B}}}-1\right)\right]\right\rbrace \tan ^{2} \chi_{0}}  & \beta=0\\
            {P_{0}\sqrt{1+\left\lbrace 1-\exp \left[-\frac{2\tau_\beta(t)}{\tau_\chi} \right] \right\rbrace\tan ^{2} \chi_{0}}} & \beta\ne0\\
            \end{array}\right.$}
	\end{equation}
    where 
    \begin{equation}
    \tau_\beta(t)=\frac{t \beta\tau_B^{2/\beta}+\tau_B^{2/\beta+1}-\tau_B(t \beta+\tau_B)^{2/\beta}}{(\beta-2)(t \beta +\tau_B)^{2/\beta}}.
    \end{equation}
    
    Based on Eq.~(\ref{dchi}), it is found that the inclination is not changed throughout the spin evolution when $\chi_{0}=0^\circ$ or $ 90^\circ$. If this is the case, Eq.~(\ref{Pt}) can be reduced to a form with no inclination evolution, which is the same form as Eq.~(\ref{eq:Pt90}) in \cite{Jawore.et.al.2020}, 
        \begin{eqnarray}
		\label{eq:Pt90}
		\lim_{\chi_{0} \to 90^\circ} P_\mathrm{vac}(t)=&&P_{0}\sqrt{1+\frac{\tau_{B}}{\tau_{\chi}}\left(1-e^{-\frac{2 t} {\tau_{B}}}\right) \tan ^{2} \chi_{0}}\\
        \lim_{\chi_{0} \to 0^\circ} P_\mathrm{vac}(t)=&& P_{0}
	\end{eqnarray}

	\subsection{Plasma-filled magnetosphere}
    If the magnetosphere of the newborn magnetar is instead plasma filled, the strong magnetic field of the magnetar would force charged particles in the wind to corotate with the star. This results in a significant loss of angular momentum which is dominant in the early stage, and will affect the spin and inclination evolution of the star \citep{Thompson.et.al.2004,Harding.et.al.1999, Thompson.et.al.2000}. This phenomenon is familiar from the study of nondegenerate stars \citep{Schatzman.1962}. Based on the findings in \cite{Lander.et.al.2020}, the loss rate of electromagnetic energy can be expressed as,
        \begin{equation}
		\label{eq:dE_EM}
		\dot{E}_{\mathrm{EM}}=\left\{\begin{array}{ll}
			c^{2} \dot{M}  \sigma_{0}^{2/3} & \sigma_{0}<1  \\
			\frac{2}{3} c^{2} \dot{M}  \sigma_{0} & \sigma_{0} \geq 1 \\
		\end{array}\right.
	\end{equation}
	where $\sigma_{0}$ is called as wind magnetization factor \citep{Ravi.et.al.2014}, and represents the ratio between the Poynting-flux and kinetic energy losses of a particle,
        \begin{equation}	
		\label{eq:sigma0}
		\sigma_{0}=\frac{4B^{2} \mathcal{F}_{\mathrm{op}}^2 R^{4} \Omega^{2}}{\dot{M}   c^{3}}.
	\end{equation}
    Here $\dot{M}$ is the mass-loss rate, which is dominated by the neutrino-driven mass-loss rate in the early stages of the newborn magnetar \citep{Metzger.et.al.2011}. 
	
	For a nonmagnetic, nonrotating star, the neutrino-driven mass-loss rate can be written as the analytical expression \citep{Qian.et.al.1996},
        \begin{equation}
		\dot{M}_{\rm \nu}=-5 \times 10^{-5} \mathrm{M}_{\odot} \mathrm{s}^{-1}\left(\frac{L_{\rm \nu}}{10^{52}~\mathrm{erg}~\mathrm{s}^{-1}}\right)^{5 / 3}\left(\frac{E_{\rm \nu}}{10~\mathrm{MeV}}\right)^{10 / 3}.
	\end{equation}
	$\dot{M}_{\rm \nu}$ is sensitively dependent on the neutrino luminosity ($L_{\rm \nu}$) and the neutrino energy ($E_{\rm \nu}$). We adopt the expressions for $L_{\rm \nu}$ and $E_{\rm \nu}$ from \citep{Lander.et.al.2020} which were obtained by fitting the simulations in \citep{Pons.et.al.1999},
        \begin{equation}
		\frac{L_{\rm \nu}(t)}{10^{52} \mathrm{erg}~\mathrm{s}^{-1}} \approx 0.7 \exp \left(-\frac{t}{1.5 \mathrm{~s}}\right)+0.3\left(1-\frac{t}{50 \mathrm{~s}}\right)^{4},
	\end{equation}
        \begin{equation}
		\frac{E_{\rm \nu}(t)}{10~\mathrm{MeV}} \approx 0.3 \exp \left(-\frac{t}{4 \mathrm{~s}}\right)+1-\frac{t}{60 \mathrm{~s}}.
	\end{equation}
	However, newborn magnetars are typically regarded as rapidly rotating, strongly magnetized NSs. The powerful magnetic field traps charged particles within the closed field lines; thus, only a fraction ($\mathcal{F}_{\mathrm {op}}$) of the particles in the open field line region has a chance to escape the star,
        \begin{equation}
		\label{eq:Fop}
		\mathcal{F}_{\mathrm {op}}=1-\sqrt{1-R/ R_{Y}},
	\end{equation}
	where $R_Y$ is the maximum distance that the magnetic field line can remain closed. On the other hand, the centrifugal force will also affect escaped particles, 
        \begin{eqnarray}
            \label{eq:Fcent}
            \mathcal{F}_{\mathrm {cent }}=&\exp \left[\left(0.33 \Omega \max \left\{R / R_{Y}, \sin \chi\right\}\right) ^{1.5}\right] \nonumber\\
            &\times\left[1-\exp \left(-R_{\mathrm{A}} / R_{\mathrm{s}}\right)\right]+\exp \left(-R_{\mathrm{A}} / R_{\mathrm{s}}\right),
	\end{eqnarray}
	where $R_\mathrm{A}$ is the $\text{Alfvén}$ radius, and $R_s=\left( GM/\Omega^2\right) ^{1/3}$ is the corotation radius where the centrifugal force is equal to gravity. The centrifugal force of rapid rotation will enhance the mass-loss rate by a factor $\mathcal{F}_{\mathrm {cent }}$ if the half-angle of the open field line ($\theta_\mathrm{op}=\arcsin \left(\sqrt{R/R_Y}\right) $) is close to the rotating equator. We adopt the same diagnostic as \cite{Lander.et.al.2020}, such as calculating the centrifugal enhancement if $\left(\chi+\theta_\mathrm{op}\right) >\pi/4$ and setting $\mathcal{F}_{\mathrm {cent }}=1$ if $\left(\chi+\theta_\mathrm{op}\right)<\pi/4$. Therefore, the realistic mass-loss rate $\dot{M}$, taking into account the effects of rotation and magnetic field, should be modified as \cite{Metzger.et.al.2011}
        \begin{equation}
		\label{eq:dM}
		\dot{M}=\dot{M}_{v} \mathcal{F}_{\mathrm {op }} \mathcal{F}_{\mathrm {cent }}.
	\end{equation}
	
	In order to obtain the realistic mass-loss rate, we follow the method of \cite{Lander.et.al.2020} to ignore the centrifugal enhancement (i.e., set $\mathcal{F}_{\mathrm {cent }}=1$). Then, we combine that with Eqs.~(\ref{eq:dM}), (\ref{eq:sigma0}), and (\ref{eq:Fop}), and employ a phenomenological relation $R_Y=R_L/\max \left\{\left(0.3 \sigma_{0}^{0.15}\right)^{-1}, 1\right\}$ \citep{Bucciantini.et.al.2006,Metzger.et.al.2007} to eliminate $\sigma_{0}$, and get
        \begin{equation}
		\left(1-\sqrt{1-\frac{R}{R_{Y}}}\right) \left(\frac{0.3R_{L}}{R_Y}\right)^{1 / 0.15}=\frac{ c^{3} \dot{M}_{v}}{4B^{2} R^{4} \Omega^{2}},
	\end{equation}
	where $R_L=c/\Omega$ is the light cylinder radius. A real solution for $R_Y$ in this equation can be obtained if $R_Y>R$. The limiting value ($R_{Y}=R$) given by \cite{Lander.et.al.2020} can be used as the result for all of the opening magnetic field ($\mathcal{F}_{\mathrm{op}}=1$) when $R_Y<R$. We adopt $R_{Y}=R$ to calculate $\sigma_0$ in Eq.~(\ref{eq:sigma0}) with the absence of centrifugal enhancement. We can estimate the $\text{Alfvén}$ radius $R_A$ by invoking both $\sigma_0$ and another phenomenological relation $R_A=R_L/\max \left\{\sigma_{0}^{-1 / 3}, 1\right\}$ given in \cite{Metzger.et.al.2011}, and calculate the centrifugal enhancement $\mathcal{F}_{\mathrm{cent}}$ by using Eq.~(\ref{eq:Fcent}). We adopt this to calculate Eq.~(\ref{eq:sigma0}) again to obtain the final magnetization $\sigma_0$. Therefore, one can write the spin evolution as
        \begin{equation}
		\label{eq:dOmegaEM}
		\dot{\Omega}_\mathrm{EM}(t)=\frac{\dot{E}_{\mathrm{EM}}(t)}{I\Omega(t)}.
	\end{equation}
	
	Reference \cite{Lander.et.al.2020} pointed out that viscosity may be a significant influence on the inclination evolution. In our calculations, we treat the star as a rigid body and ignore the contribution from internal viscosity to the inclination evolution,
        \begin{equation}
		\label{eq:dChi}
		\dot{\chi}_\mathrm{EM}(t)=\frac{\dot{E}_{\mathrm{EM}}\sin\chi(t)\cos\chi(t)}{I \Omega^{2}}.
	\end{equation}
	Both $L_{\rm \nu}$ and $E_{\rm \nu}$ decay rapidly and become negligible during the first 1 min when the star gradually becomes neutrino transparent. Therefore, we simply employ the first 40 s of spin and inclination evolution in Eqs.~(\ref{eq:dE_EM}) and (\ref{eq:dChi}), respectively \citep{Lander.et.al.2020}. 

	After 40 s, the magnetar wind weakens significantly and magnetospheric torque begins to dominate the evolution. We adopt the expression given by \cite{Spitkovsky.et.al.2006} and \cite{Philippov.et.al.2014}. It is obtained from the simulation of a pulsar magnetosphere that depicts the development of the spin,
        \begin{equation}
		\label{dOmega2}
		\dot{\Omega}_\mathrm{pla}(t)=-\frac{R^{6}}{ Ic^{3}}  B(t)^{2}\left(k_0+k_1 \sin ^{2} \chi(t) \right)\Omega(t)^{3},
	\end{equation}
	and the inclination evolution,
        \begin{equation}
		\dot{\chi}_\mathrm{pla}(t)=-k_2 \frac{R^{6}}{I c^{3}} B(t)^{2} \sin \chi(t) \cos \chi(t) \Omega(t)^{2}.
	\end{equation}
	Here we adopt the approximation $k_0\approx k_1 \approx k_2 \approx 1$ given by \cite{Philippov.et.al.2014} to perform the calculations\footnote{The exact values of $k_0$ and $k_1$ remain subject to debate, with a possible variation on the order of about 20-40\%; however, the effect of this difference does not alter our overall conclusion.}, and thus the relationship between inclination and angular frequency is established as
        \begin{equation}
		\label{sin^2_chi}
		\sin^2 \chi=\left( \frac{\sqrt{4 \Omega^{2} \sin ^{2} \chi_{0}+\Omega_{0}^{2} \cos ^{4} \chi_{0}}-\Omega_{0} \cos ^{2} \chi_{0}}{2 \Omega \sin \chi_{0}}\right) ^2.
	\end{equation}
	Then, we can rewrite Eq. (\ref{dOmega2}) by using (\ref{sin^2_chi}), and present the spin evolution as the derivative of the period,
        \begin{equation}
		\label{eq:dp2}
		\dot{P}_\mathrm{pla}(t)=\frac{4 \pi^2 R^6 B(t)^2}{c^3 I P(t)}\left[1+\left(\sqrt{1+P(t)^2 C_0^2}-P(t) C_0\right)^2\right],
    \end{equation}
	where
        \begin{equation}
		C_0=\frac{\cos \chi_0  ^2}{2 P_0 \sin \chi_0}
	\end{equation}
	is a constant. Notice that Eq.~(\ref{eq:dp2}) is performed after 40 s and the values of $P_0$ and $\chi_0$ used here are taken from the solutions of Eqs.~(\ref{eq:dOmegaEM}) and (\ref{eq:dChi}). The solution of Eq.~(\ref{eq:dp2}) can be expressed as
        \begin{equation}
		\label{eq:Pt2}
		P_\mathrm{pla}(t)=f^{-1}\left(x \right),
	\end{equation}
        \begin{equation}
		f(x)=\frac{C_0 x (C_0 x + \sqrt{1 + C_0^2 x^2})-\mathrm{arcsinh}(C_0 x)}{2 C_0^2},
	\end{equation}
	where $x$ is a time-dependent variable
        \begin{widetext}
        \begin{eqnarray} 
        x(t)=\left\{\begin{array}{ll}
            \frac{P_0^2}{2} \left[ 1+\frac{\sqrt{1+ C_0^2 P_0^2}}{P_0 C_0}-\frac{\mathrm{arcsinh}(C_0 P_0)}{P_0^2 C_0^2}+\frac{3 \left( 1 - \exp(-2 t/\tau_B) \right) \tau_B}{\tau_{\chi}\cos\chi_0^2}\right], & \beta=0;\\
            \frac{P_0^2}{2} \left[ 1+\frac{\sqrt{1+ C_0^2 P_0^2}}{P_0 C_0}-\frac{\mathrm{arcsinh}(C_0 P_0)}{P_0^2 C_0^2} +\frac{6 \tau_\beta(t)}{\tau_{\chi}\cos\chi_0^2} \right], & \beta\ne0.\\
            \end{array}\right.
	\end{eqnarray}
        \end{widetext}

\section{Sample selection and the statistical analysis of magnetars}
\begin{figure*}
    \includegraphics[width=3.5in]{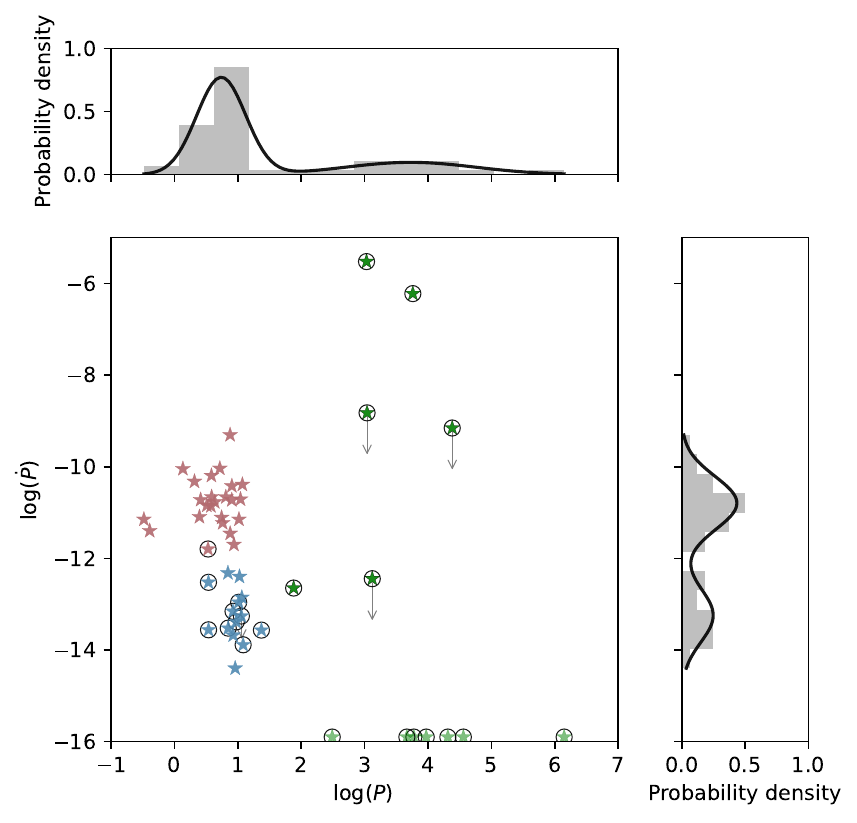}
    \caption{The spin period $P$ against period derivative $\dot{P}$ for the magnetar and candidates. The data of candidates (circled) collected by \cite{Beniamini.et.al.2023}. In addition, a recently discovered long-period radio source GPM J1839-10 is also included \citep{Hurley.et.al.2023}. The statistical distribution of $P$ (top panel) shows a bimodal distribution with a separated line at $P\sim$68 sec, dividing the population into “short-P” (red and blue stars) and “long-P” subclasses (green stars). The statistical distribution of $\dot{P}$ (right panel) for only the "short-P" subclass also shows a bimodal distribution with a dividing line at $\dot{P}\sim 7.5\times 10^{-13}$. The red star indicates the “high-$\dot{P}$ short-P” subclass, and the green star indicates the “low-$\dot{P}$ short-P” subclass of magnetars. The transparent green stars at the bottom indicate objects whose $P$ are observed only. }
    \label{Fig1:P_Pdot}
\end{figure*}

    The known magnetars from observations that we adopted in this paper are taken from the McGill catalog \citep{Olausen.et.al.2014}, and the magnetar candidates are collected by \cite{Hurley.et.al.2022,Beniamini.et.al.2023}. In addition, the recently discovered long-period radio source GPM J1839-10 \citep{Hurley.et.al.2023} and the longest period (e.g., 6.67 h) source 1E161348-5055 in supernova remnant RCW 103 \citep{DeLuca.et.al.2006} are also included in our sample. In Fig.\ref{Fig1:P_Pdot}, we plot the period derivative as a function of the spin period for magnetars and candidates ($P-\dot{P}$ diagram).
    
    In the statistical analysis of $P$ and $\dot{P}$ for all magnetars or candidates, it is found that the periods of all magnetars or candidates appear a bimodal distribution with a separation line at $P\sim$68 s. We define them as “long-P” for period $>68$ s (green stars in Fig. \ref{Fig1:P_Pdot}) and “short-P” for period $<68$ s (red and blue stars in Fig. \ref{Fig1:P_Pdot}) magnetar subclasses, respectively. If we believe that the bimodal distribution of the period exists intrinsically, it hints that magnetars may not form a homogenous class, possibly tracking different evolution channels for different subclasses. There may be other mechanisms driving the spin evolution channel for the “long-period” magnetar subclass, such as the accretion torques that are dominant in 4U 0114+65 \citep{Torrej.et.al.2018} and SXP 1062 \citep{Haberl.et.al.2012}. On the other hand, it is found that all of the “long-period” subclass are magnetar candidates do not have a measured $\dot{P}$ value excepting three cases (see Fig. \ref{Fig1:P_Pdot} and Table \ref{tab:observation}). We focus our investigations in this paper only on the ``short period'' magnetar subclass due to the lack of measurements of ${\dot P}$ for the “long-period” subclass and the fact that they are magnetar candidates.

    For the “short-P” magnetar subclass, we find that the $\dot{P}$ values also form a bimodal distribution with a dividing line at $\dot{P}\sim 7.5\times 10^{-13}$ (see Fig. \ref{Fig1:P_Pdot}). We have named the two regions of the distribution “high-$\dot{P}$ short-P” and “low-$\dot{P}$ short-P” magnetar subclasses. It seems that the evolution of the magnetars does not share similar channels even though their periods are all short, if the bimodal distribution of $\dot{P}$ is intrinsic. “high-$\dot{P}$” and “low-$\dot{P}$" indicate stronger and weaker inferred dipole fields, respectively. The weakest inferred field in our sample is from a known magnetar SGR 0418+5729. However, the dipole component that we observe is only a projection of the true magnetic field \cite{Braithwaite.2009, Gourgouliatos.2018}, and it is possible that it has a toroidal component with a higher order via crustal Hall evolution \cite{Wood.Hollerbach.2015, Gourgouliatos.et.al.2016}. This may imply that magnetar classification is related to the configuration of the magnetic field. 
    Moreover, one needs to clarify whether or not the bimodal distribution of both $P$ and $\dot{P}$ is possibly caused by selection effects; namely, the number of observed objects may not be large enough to result in a true bimodal distribution.

\section{Details of Monte Carlo Simulations}
\subsection{Monte Carlo simulation procedures}
    In order to investigate the effect of varying inclination on the magnetar spin evolution in different environments, we follow a similar Monte Carlo method to the one given in \cite{Jawore.et.al.2020} to generate synthetic magnetar populations. In this section, we present the details of these Monte Carlo simulations.
 
   First, we generate 100 synthetic newborn magnetars with zero age, where the initial spin period $P_{0}$ and surface magnetic field $B_{0}$ of the magnetar population are random variables following the log-normal distribution with $(\mu_{P_{0}},\sigma_{P_{0}})$ and $(\mu_{B_{0}},\sigma_{B_{0}})$, respectively \citep{Johnson.et.al.1995}. $\mu$ and $\sigma$ represent the expected value and standard deviation of the log-normal distribution, respectively. The lifetime of each magnetar within the synthetic population is another parameter that must be specified. We assume that the lifetime of each magnetar obeys a uniform density distribution $t\sim \mathrm{U}(0,t_\mathrm{max})$, where $t_\mathrm{max}$ represents the maximum lifetime of the synthetic magnetar populations. From an observational point of view, the lifetime of a magnetar is typically less than 100 kyr, or even younger (see Table.~\ref{tab:observation}). In our simulations, the maximum evolution time is fixed to $t_\mathrm{max}=50$ kyr.
   
    Second, by considering the evolution of the magnetic field in Eq.~(\ref{Bt}), a fraction of magnetars may be nondetectable before they evolve to $t_\mathrm{max}$, because the majority of the magnetic energy has gradually waned. This process is called ``fade-away" and is described in \cite{Jawore.et.al.2020}. One can define the probability function of the ``fade-away" which is related to the strength of the magnetic field as follows:
        \begin{equation}
		P_{\mathrm {fade }}(B) = 1-\left[1+\left(\frac{s_{1}}{B}\right)^{s_{2}}\right]^{-1}
    \end{equation}
   Here, $s_2=1.79$ is adopted from \cite{Jawore.et.al.2020}, while $s_1$ is a free parameter obtained by fitting the magnetic field of the observations in Table.~\ref{tab:observation} via the empirical cumulative distribution function. If this is the case, due to the ``fade-away", one needs to know what fraction of newborn magnetars can be detected as the magnetic field decays with time. Based on the results in \cite{Beniamini.et.al.2019}, the formation rate of galactic magnetars is $(2.3-20)~\mathrm{kyr}^{-1}$ without considering the decay of the magnetic field. However, the birth rate (BR) of galactic magnetars is as high as $20~\mathrm{kyr}^{-1}$ if one considers the decay of the magnetic field \citep{Colpi.et.al.2000,Keane.et.al.2008}. In our simulations, we adopt $20~\mathrm{kyr}^{-1}$ as the magnetar birth rate upper limit and try to search for the optimal parameter distribution. Here, we eliminate those parameters which have exceeded this limit in the magnetar population. Because of the poorly constrained magnetar equation of state, we fix $M=1.4 ~M_{\odot}$, $R=10^6$ cm, and $I=1.11\times10^{45}~\mathrm{g~cm}^2$ in our simulations. The other parameters related to model evolution can be represented as a vector$\overrightarrow{\theta}$,
 
    \begin{equation}
        \scalebox{0.85}{$\overrightarrow{\theta}=\left\{
            \begin{array}{ll}            \left(\mu_{P_{0}},\sigma_{P_{0}},\mu_{B_{0}},\sigma_{B_{0}},\tau_B,s_1, \beta\right), & \mathrm{without~inclination~evolution}\\               \left(\mu_{P_{0}},\sigma_{P_{0}},\mu_{B_{0}},\sigma_{B_{0}},\chi_0,\tau_B,s_1,\beta\right), & \mathrm{with~inclination~evolution}\\  
        \end{array}\right.$}  
    \end{equation}
   
     Given the initial values of those parameters and following Eqs.~(\ref{dP}), (\ref{Pt}), (\ref{eq:dp2}), and (\ref{eq:Pt2}) evolution, one can obtain the distribution of synthetic magnetars in their present ages on the $P-\dot{P}$ diagram.

	Third, in order to quantitatively describe the goodness of fit between synthetic and observable magnetar populations, we performed two independent Kolmogorov-Smirnov (KS) tests for $P$ and $\dot{P}$, respectively. The properties of observed magnetars are listed in Table \ref{tab:observation}.
	Following the method of \cite{Ridley.et.al.2010} and \cite{Jawore.et.al.2020}, we adopt a figure of merit (FOM) as an indicator of fit goodness,
    \begin{equation}
		\mathrm { FOM }=(1-p_{\rm KS}(P))+(1-p_{\rm KS}(\dot{P}))
    \end{equation}
	where $p_{\rm KS}(P)$ and $p_{\rm KS}(\dot{P})$ are the KS test $p$ value for $P$ and $\dot{P}$, respectively, and the FOM range is $0<\mathrm{FOM}<2$. A smaller FOM value indicates a closer match between synthetic and observational magnetar populations. When the selected $\overrightarrow{\theta}$ produces synthetic populations that are far away from the observed magnetar population, it indicates that random processes no longer have a substantial effect on FOM. If this is the case, one only needs to run the Monte Carlo simulations one more time and obtain the value of FOM. When the synthetic populations are close to the observed one, FOM will fluctuate wildly as a result of random processes. In order to avoid the risk that a single computation of FOM and BR may diverge far away from the average value, we gradually increase the number of iterations to 1000 times, and then calculate the average values of $\overline{\mathrm{FOM}}$ and $\overline{\mathrm{BR}}$.
	
	Finally, the goodness of fit is strongly dependent on the values of the selected initial parameters. In order to test the dependence of the goodness of fit on the parameters, we employ an automatic algorithm that is described in \cite{Jawore.et.al.2020} to find the best distribution of the parameters. This algorithm was first employed by \cite{Gullon.et.al.2014} via the simulated method in \cite{Press.et.al.1993}. They invoked a one-dimensional KS test to judge goodness of fit, and applied this algorithm to search for the optimal parameters. Afterwards, \cite{Jawore.et.al.2020} employed the algorithm to the FOM with two-dimensional KS tests instead of one-dimensional ones as the goodness of fit. For a more detailed description of the algorithm one can refer to \cite{Jawore.et.al.2020}. The primary steps of a cycle of the algorithm in our simulations are shown as follows,
    \begin{enumerate}[label = (\roman*).]
		\item Generate a set of random parameters $\overrightarrow{\theta_i}$ within a given range. Then, set the initial values of the temperature parameter ($T$), the maximum accepted step ($\mathrm{N_{acc}}$), and the sample size ($\mathrm{N_{pop}}$) for calculating $(\overline{\mathrm{FOM}})$.
		\item By using $\overrightarrow{\theta_i}$ to evolve a synthetic population of 100 magnetars, we calculate values of $(\mathrm{FOM})_i$ and $(\overline{\mathrm{BR}})_i$ with KS tests of the two-sample (synthetic and observational populations). If it exceeds 20 $\mathrm{kyr}^{-1}$, then return to step (ii). Repeat this process $\mathrm{N_{pop}}$ times and calculate $(\overline{\mathrm{FOM}})_i$ again.
		\item We check whether $(\overline{\mathrm{FOM}})_{i}$ is better (or smaller) than $(\overline{\mathrm{FOM}})_\mathrm{{min}}$. If so, we set $(\overline{\mathrm{FOM}})_\mathrm{{min}}=(\overline{\mathrm{FOM}})_{i}$ and save $\overrightarrow{\theta_i}$; if not, generate a random variable $R$ with a uniform distribution between 0 and 1. If $R<\exp\left[\left(\overline{\mathrm{FOM}}_\mathrm{{min}}-\overline{\mathrm{FOM}}_{i}\right)/T\right]$, then set $(\overline{\mathrm{FOM}})_\mathrm{{min}}=(\overline{\mathrm{FOM}})_{i}$ and save $\overrightarrow{\theta_i}$ as well, otherwise return to step (ii).
		\item Repeat the processes of steps (i)-(iii) until a significant low-$\overline{\mathrm{FOM}}$ area occurs, or the maximum accepted step ($\mathrm{N_{acc}}$) is reached.
    \end{enumerate}
	{\bf Caution}: $\mathrm{N_{acc}}$ must be large enough to prevent the algorithm from stopping before a possible low-$\overline{\mathrm{FOM}}$ region appears. Therefore, we set $\mathrm{N_{acc}}=10^5$ as the initial cycles. Given limited computational resources and the dependence of $\mathrm{N_{acc}}$ on the degrees of freedom in the parameter space, $\mathrm{N_{pop}}=10$ and $T= 100$ are adopted as the initial cycles, and $\mathrm{N_{pop}}$ is gradually increased to 50, while $T$ is gradually decreased to 0.1. 

    \subsection{Application for short-P magnetar subclass}
\begin{figure*}
    \includegraphics[width=7in]{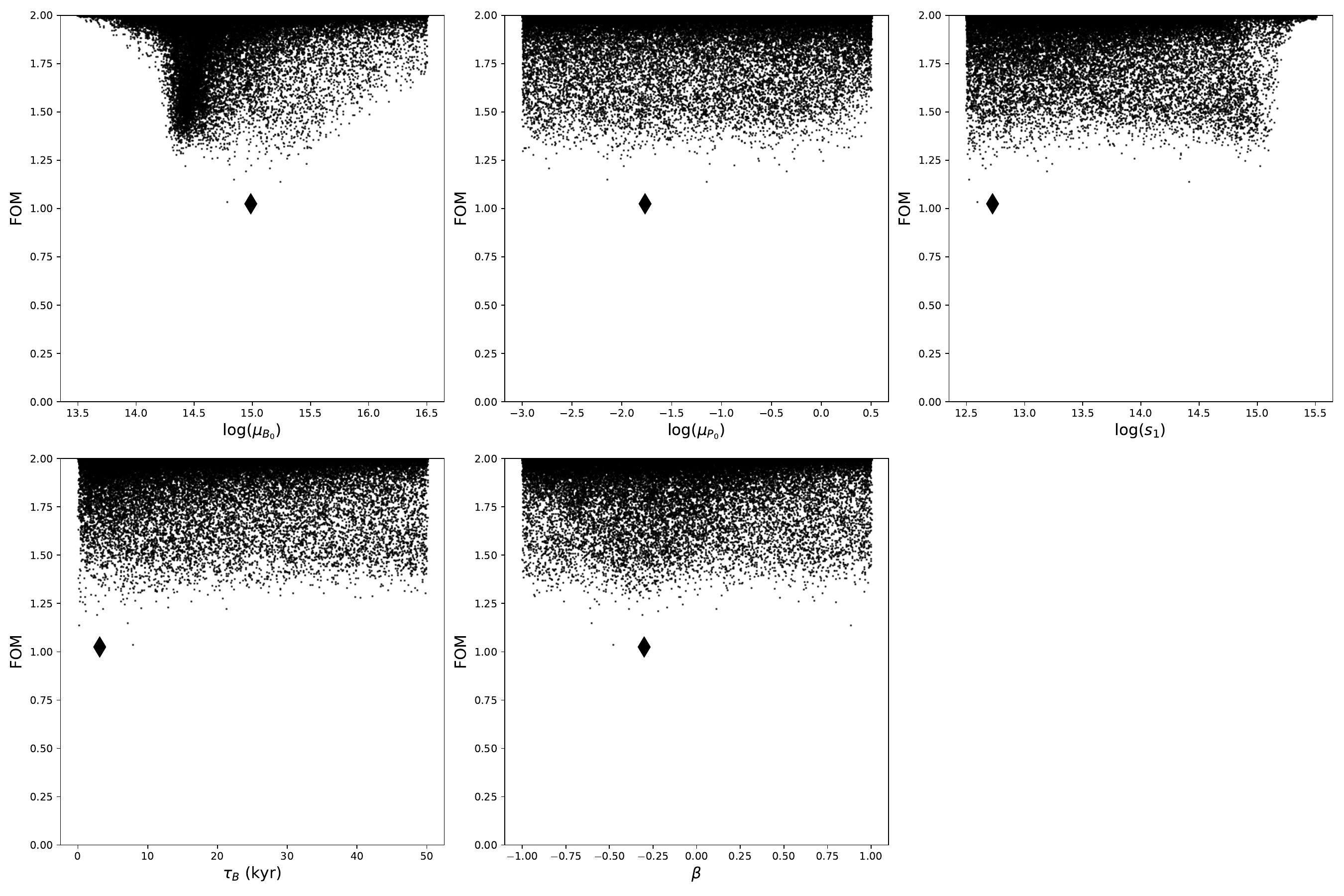}
    \caption{Values of $\overline{\mathrm{FOM}}$ for different free parameters in vacuum magnetosphere and without inclination evolution via simulation and test with the "short-P" subclass of magnetars. The black diamonds mark the minimum $\overline{\mathrm{FOM}}$. }
    \label{Fig2}
\end{figure*}

\begin{figure*}
    \includegraphics[width=3.5in]{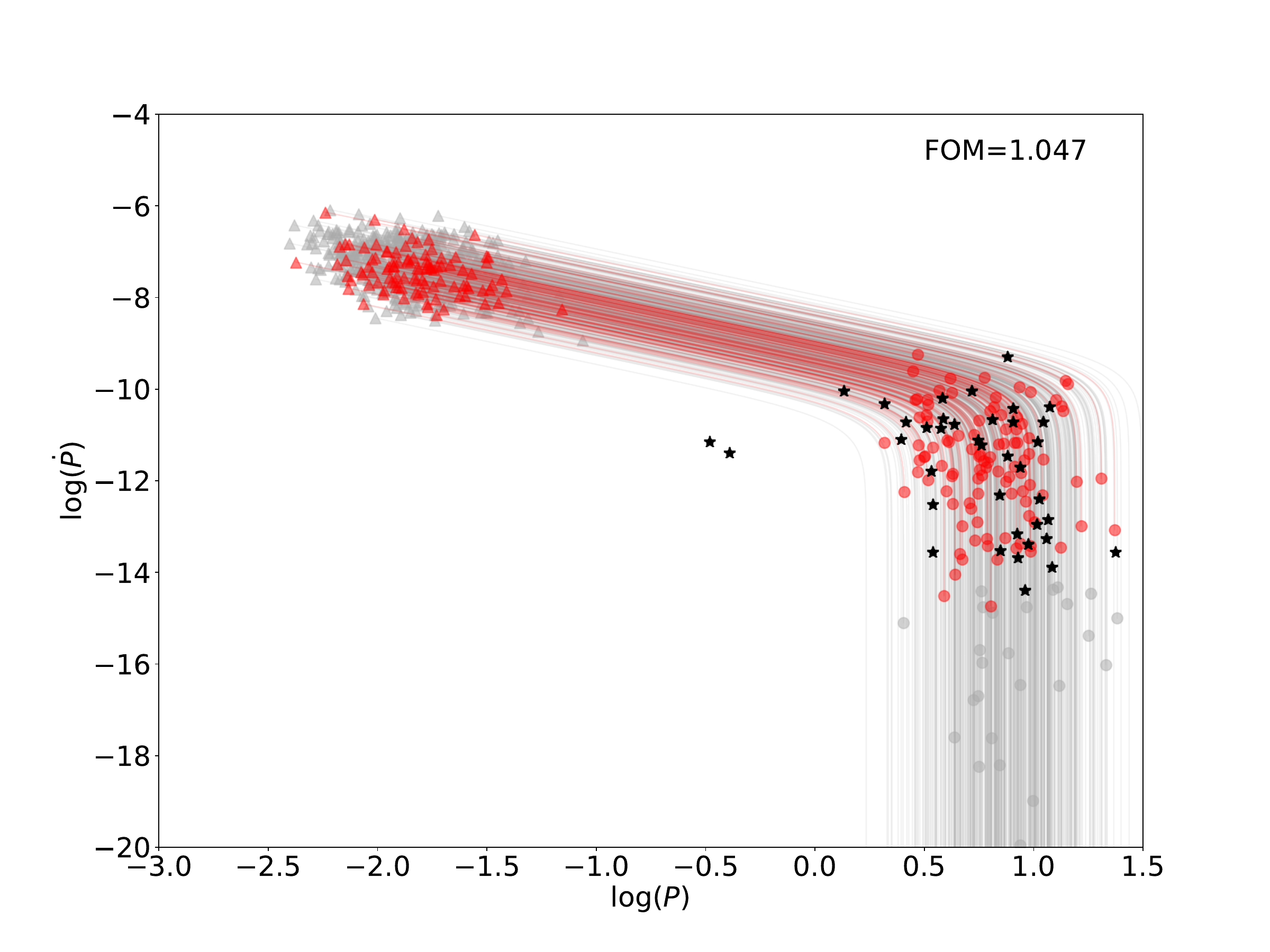}
    \caption{Evolutionary tracks of magnetar populations in vacuum magnetosphere without inclination evolution. The initial parameters correspond to the black diamonds in Fig.~\ref{Fig2}, and the black stars used for KS tests are the observed "short-P" subclass (red and blue stars of Fig.~\ref{Fig1:P_Pdot}). The red triangles and circles are the zero age and current age of magnetar populations, respectively, and their connecting lines are the evolutionary tracks. The grey dots are the fade-away population.}
    \label{Fig3}
\end{figure*}

    Based on the aforementioned method, we first apply it to the short-P magnetar subclass in a vacuum environment, adopting a fixed-inclination model. Figure \ref{Fig2} shows the values of $\overline{\mathrm{FOM}}$ for the different free parameters $\mu_{B_0}$, $\mu_{P_0}$, $s_1$, $\tau_B$, and $\beta$. It shows that no optimal set of parameters could be found to align with the distribution of the short-P population with 100,000 iterations of random walks of the free parameters. The best fit is marked by the black diamond ($\overline{\mathrm{FOM}}>1$), and the evolutionary tracks based on this set of best-fit parameters are plotted in Fig.~\ref{Fig3}. The synthetic populations seem to be able to mix with the observed samples, but their distribution performs inadequately in KS tests. It hints that the short-P magnetar subclass may not be a homogenous class. Together with the bimodal distribution of $\dot{P}$ for the short-P magnetar subclass, this strongly suggests that they may be tracking different evolution channels for high-$\dot{P}$ and low-$\dot{P}$ of short-P magnetars, respectively. It is natural in explaining why we cannot find optimal parameters for the short-P population via the simulations. Below, we will focus only on studying high-$\dot{P}$ and low-$\dot{P}$ short-P magnetars. It should be noted that the poor fitting of "short-P" magnetars may be related to the evolution model. Therefore, we perform the simulations by adopting four evolution models: with (without) inclination evolution in vacuum, and with (without) inclination evolution in plasma-filled magnetosphere, respectively.
    
    \section{Simulation of Magnetar Populations Evolution In the $P-\dot{P}$ Diagram}
      In this section, we calculate the spin evolution of different magnetar class simulations by utilizing different evolution models. We adopt "high-$\dot{P}$ short-P" and "low-$\dot{P}$ short-P" subclasses as criteria to find out how sensitively the model depends on the free parameters, and test the $P-\dot{P}$ distribution of the synthetic population to compare with observations.
      
    \subsection{Vacuum magnetosphere simulation}
\begin{figure*}
    \includegraphics[width=7in]{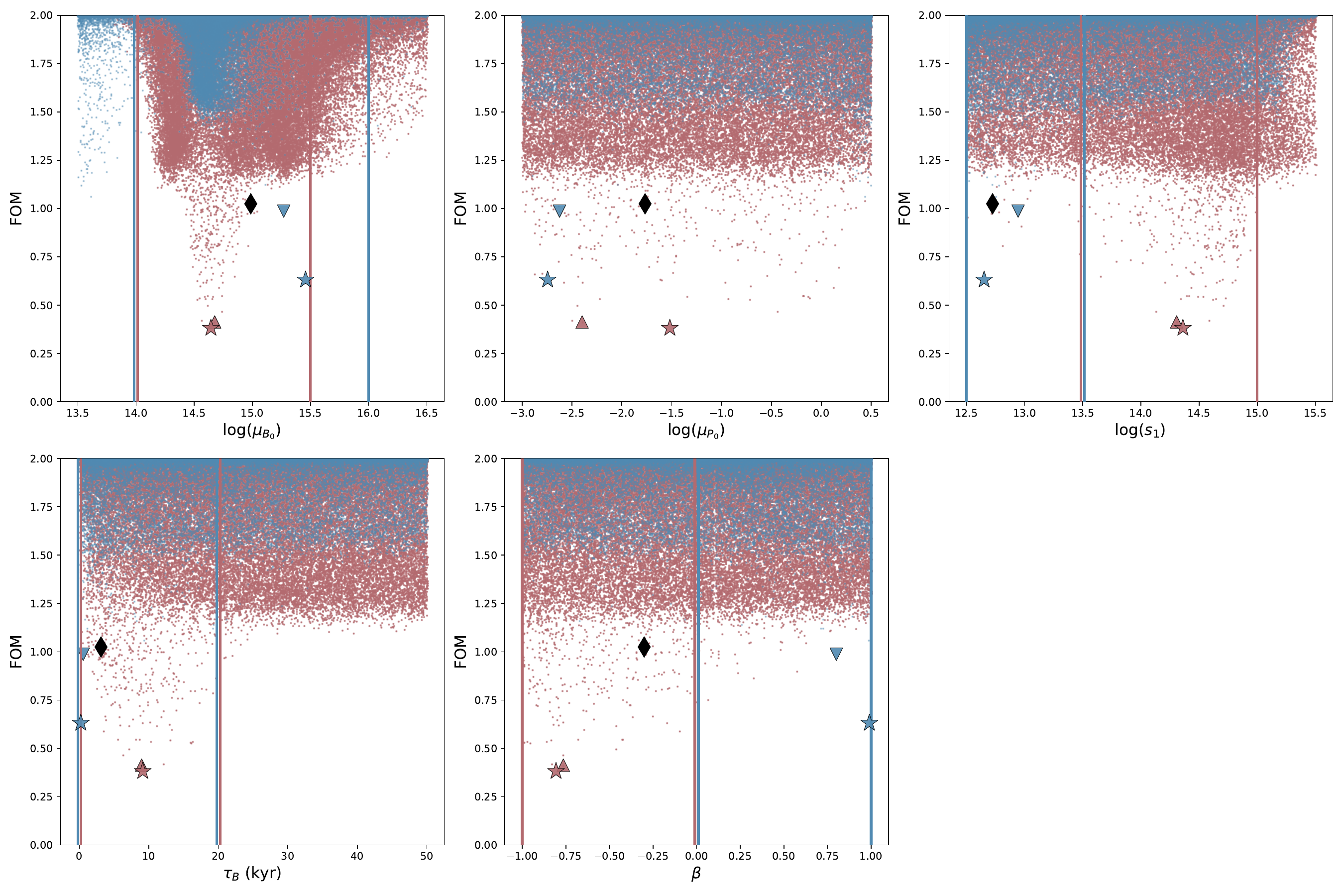}
    \caption{Values of $\overline{\mathrm{FOM}}$ for different free parameters in vacuum magnetosphere via first cycle simulation without inclination evolution. The red dots represent the fitting results of the synthetic magnetars that test with the “high-$\dot{P}$ short-P” subclass of magnetars, and the blue dots correspond to the results that test with the “low-$\dot{P}$ short-P” subclass of magnetars. The black diamonds are the best-fit results with the “short-P” subclass, same as the ones in Fig.~\ref{Fig2}. The red triangles and blue upside-down triangles represent the minimum $\overline{\mathrm{FOM}}$ in the first cycle of the "high-$\dot{P}$ short-P" and "low-$\dot{P}$ short-P" subclasses, respectively. The range of two vertical lines of the same color is the limit of parameters used for the second cycle. The respective results of the final cycle are marked as red and blue stars.}
    \label{Fig4}
\end{figure*}

\begin{figure*}
    \includegraphics[width=7in]{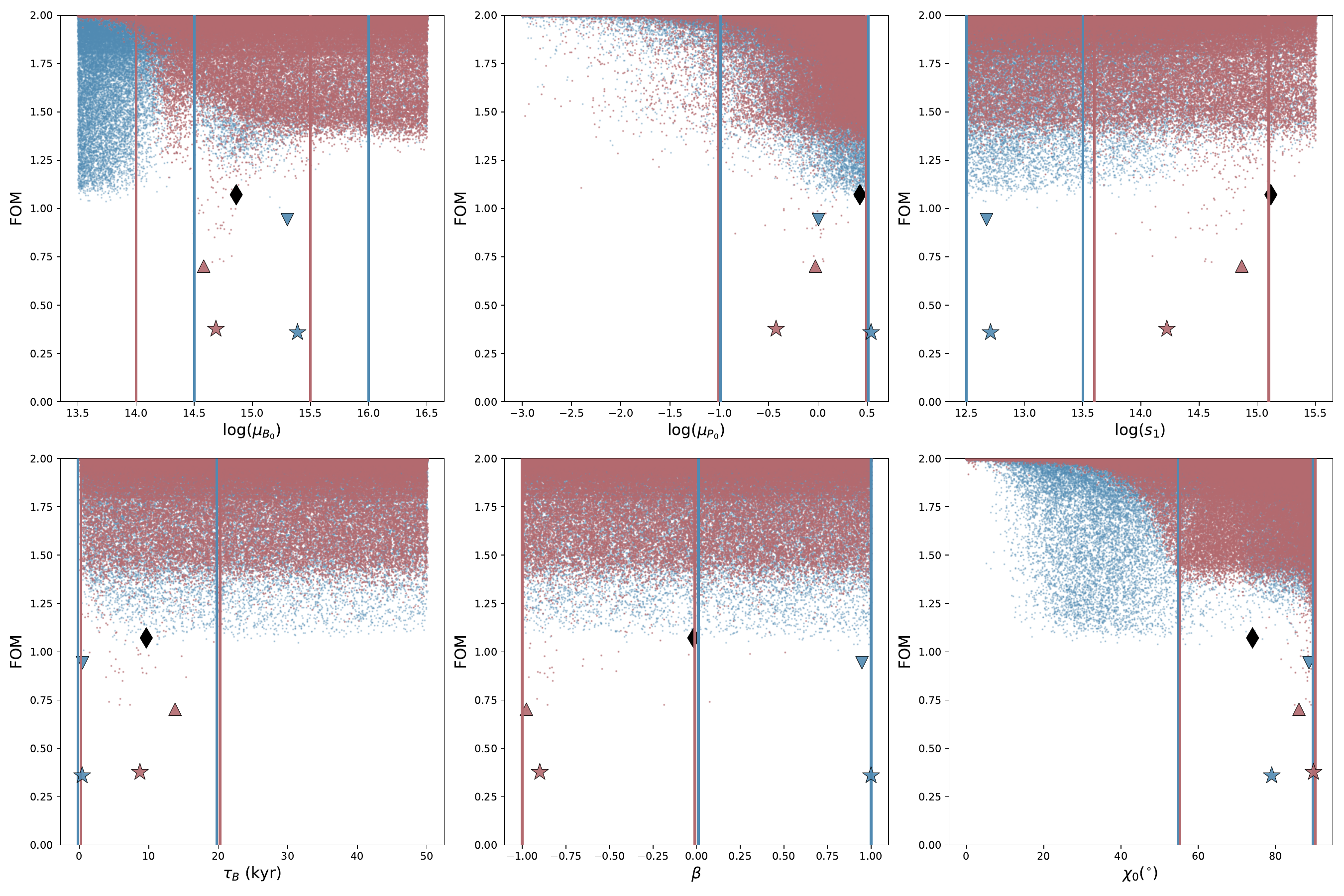}	
    \caption{Similar to Fig.~\ref{Fig4}, but considering the evolution of inclination.}
    \label{Fig5}
\end{figure*}

    \begin{itemize}	
        \item[$\bullet$] Within the scenario of no inclination evolution (i.e., the orthogonal rotators model with $\chi_0=90^{\circ} $), we simulate the magnetar evolution and run an automatic algorithm by using Eqs.~(\ref{dP}) and (\ref{Pt}). 
        Figure \ref{Fig4} shows the values of $\overline{\mathrm{FOM}}$ for the different free parameters $\mu_{B_0}$, $\mu_{P_0}$, $s_1$, $\beta$, and $\tau_B$. The minimum values of $\overline{\mathrm{FOM}}$ for the "high-$\dot{P}$ short-P" and "low-$\dot{P}$ short-P" subclasses in the first cycle are marked as the triangles and upside-down triangles, respectively. It is found that the fitting results are much improved after separately considering "high-$\dot{P}$ short-P" and "low-$\dot{P}$ short-P" subclasses from the short-P subclass. The distributions of both subclasses (high-$\dot{P}$ and low-$\dot{P}$) are dependent on $\mu_{B_0}$, $s_1$, $\beta$, and $\tau_B$. However, the low-$\overline{\mathrm{FOM}}$ regions (the range of the vertical line of the same color in Fig.~\ref{Fig4}) are obviously different for some parameters.

        On the other hand, one finds that the "high-$\dot{P}$ short-P" magnetar subclass has higher $s_1$ and $\tau_B$, but smaller $\mu_{B_0}$ than the "low-$\dot{P}$ short-P" subclass. The $\beta$ of both subclasses has a symbolic difference, but neither of them is significantly dependent on $\mu_{P_0}$. Therefore, we gradually reduce the range of other parameters within red and blue vertical lines in Fig.~\ref{Fig4} for the "high-$\dot{P}$ short-P" and "low-$\dot{P}$ short-P" subclasses, respectively. Then, we  run the next cycle until the low-$\overline{\mathrm{FOM}}$ region is no longer seen. If this is the case, we think the algorithm has converged at this cycle. The red and blue stars in Fig.~\ref{Fig4} mark the respective minimum $\overline{\mathrm{FOM}}$ of those parameters for the final cycle. The values of corresponding parameters are presented in Table.~\ref{tab:parameter}. In order to compare the results from the "high-$\dot{P}$ short-P" and "low-$\dot{P}$ short-P" subclasses with that of the short-P subclass, we also plot the result from the short-P subclass in Fig.~\ref{Fig4} marked as a diamond.
        
        \item[$\bullet$] Within the scenario of inclination evolution, more free parameters are considered in the simulations. Figure \ref{Fig5} is similar to Fig.~\ref{Fig4}, but one more parameter ($\chi_{0}$) is added. For comparison, we also present the best-fit result (black diamond in Fig.~\ref{Fig5}) for short-P magnetars. It is found that the results exhibit a similarly poor fitting to the case with inclination evolution. In contrast, improved outcomes can be realized by dividing the short-period subclass into "high-$\dot{P}$ short-P" and "low-$\dot{P}$ short-P" subclasses separately. Both distributions of the "high-$\dot{P}$ short-P" and "low-$\dot{P}$ short-P" subclasses are dependent on $\mu_{B_0}$, $s_1$, $\tau_B$, and $\beta$ within low-$\overline{\mathrm{FOM}}$ regions, and this is similar to that of the scenario without inclination evolution. However, the model is significantly dependent on both $\mu_{P_0}$ and $\chi_0$. We rerun the automatic algorithm and present the respective final results as marked stars. The corresponding results are also presented in Table.~\ref{tab:parameter}.
    \end{itemize}
    
\begin{figure*}
    \includegraphics[width=3.0in]{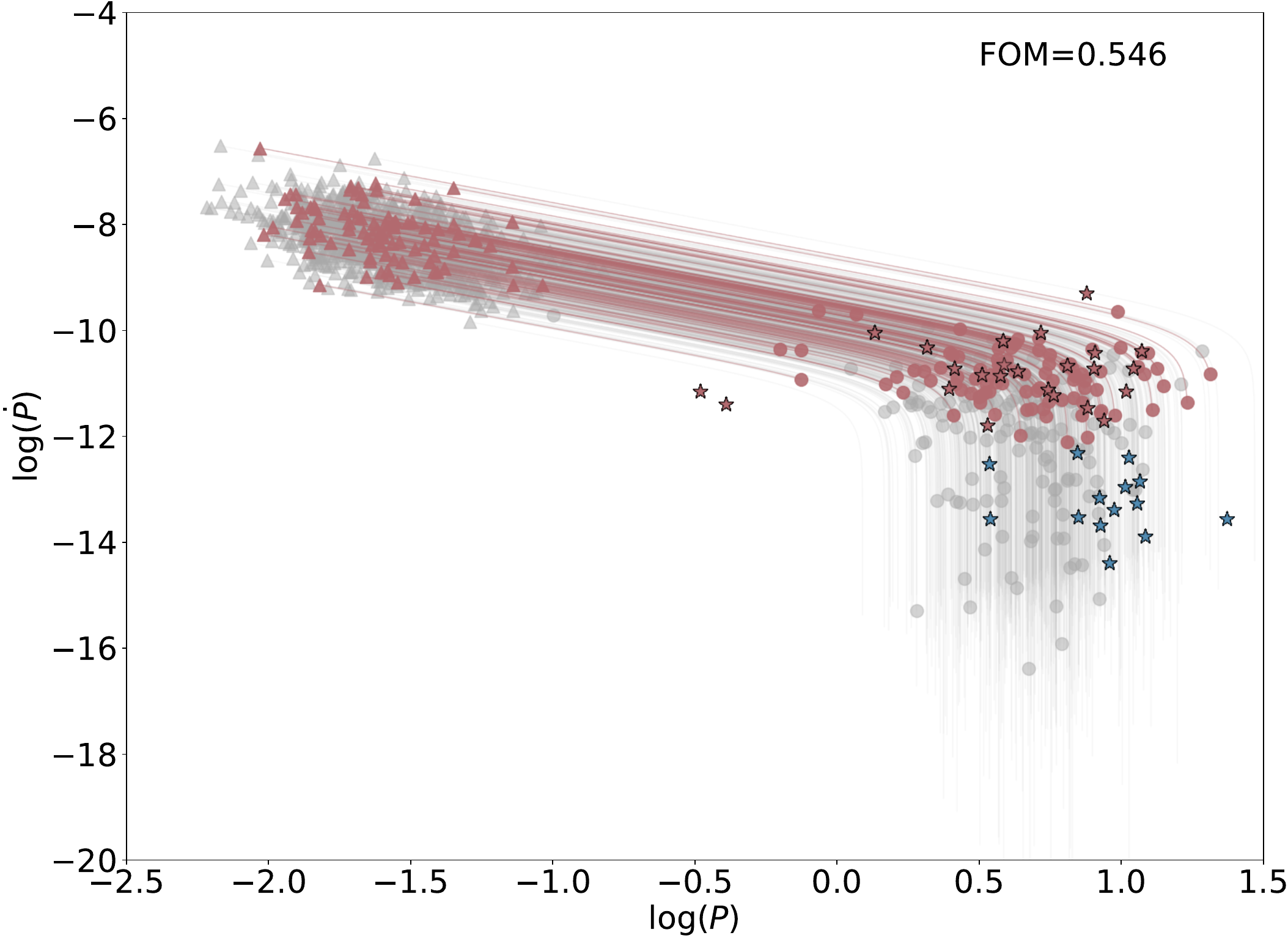}
    \includegraphics[width=3.0in]{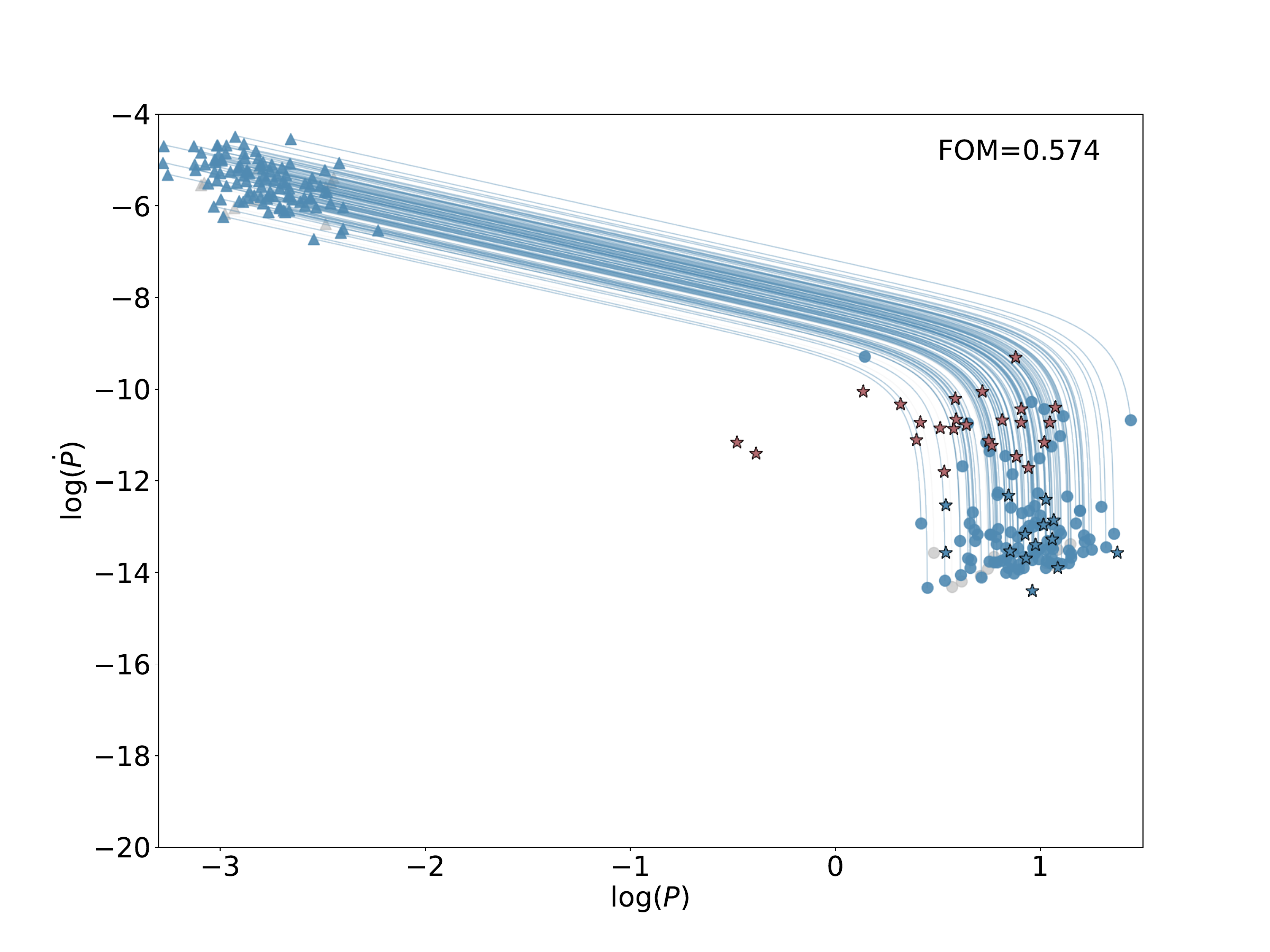}\\
    \includegraphics[width=3.0in]{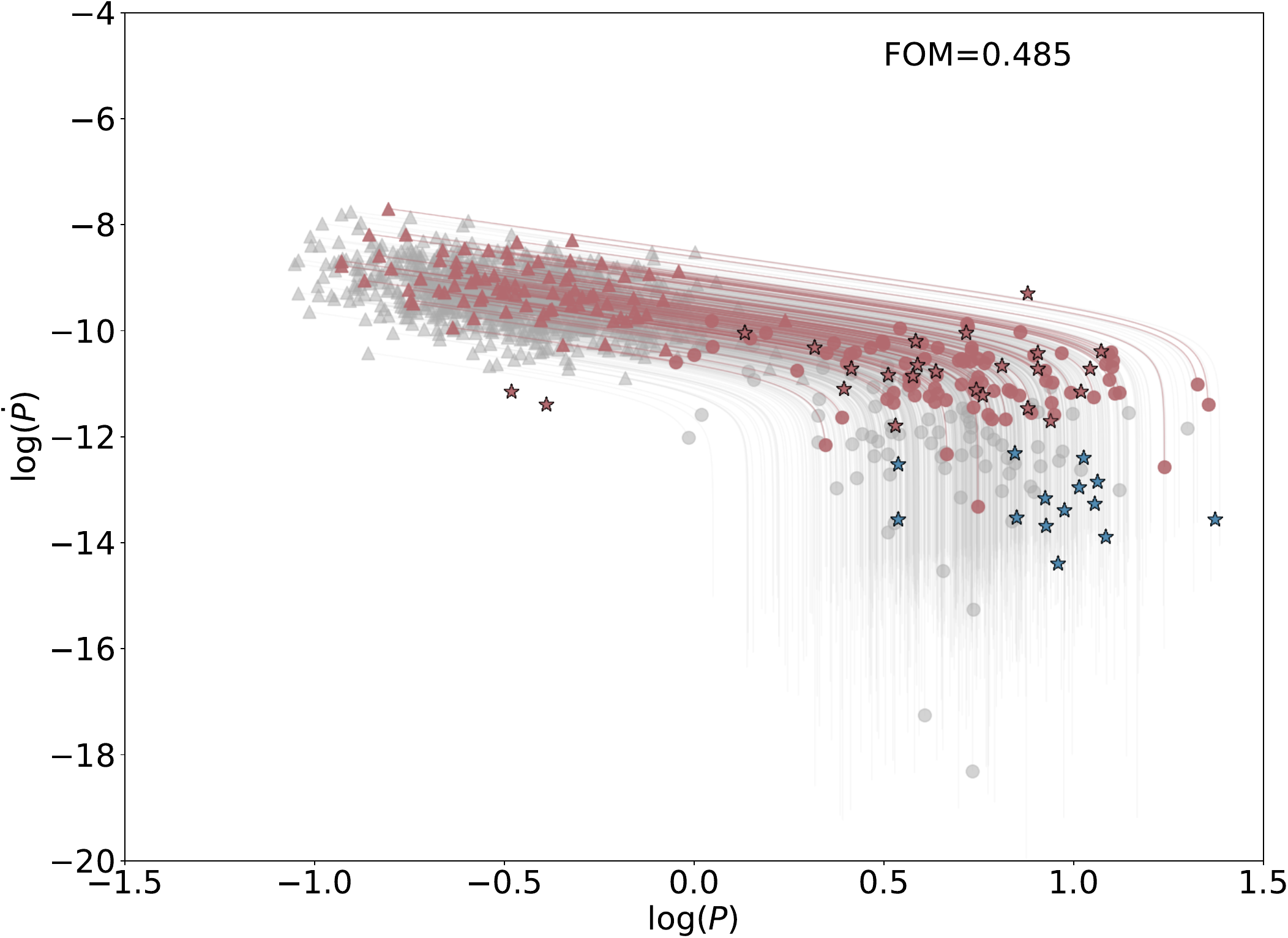}
    \includegraphics[width=3.0in]{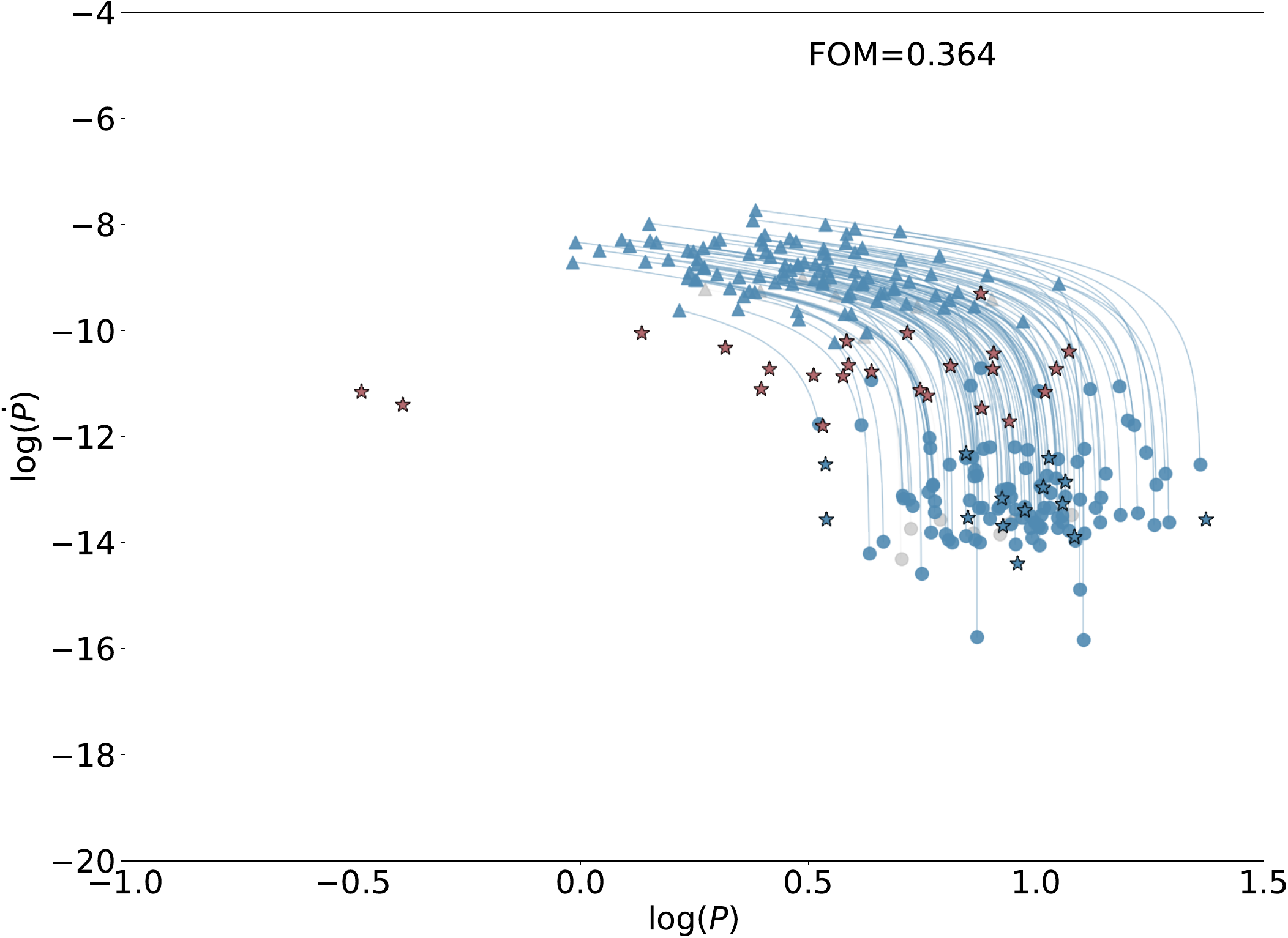}
    \caption{The evolutionary tracks of the "high-$\dot{P}$ short-P" (left panel) and "low-$\dot{P}$ short-P" subclass (right panel) of magnetars in vacuum magnetosphere without(top panel) and with(bottom panel) inclination evolution by using the parameters in Table ~\ref{tab:parameter}.}
    \label{Fig6}
\end{figure*}

     By adopting the above best-fit parameters in Table \ref{tab:parameter}, we simulate the evolution tracks from synthetic magnetar populations to the observations of "high-$\dot{P}$ short-P" and "low-$\dot{P}$ short-P" subclasses, respectively (see Fig.~\ref{Fig6}). For the "high-$\dot{P}$ short-P" subclass, we find that the population of simulated magnetars is mixed with that of observations excepting the fade-away population, and the values of $\overline{\mathrm{FOM}}$ with or without inclination evolution are as low as about 0.5 (see Table \ref{tab:parameter}). This means that the simulation results seem consistent with the observed magnetar population whether or not we consider the inclination evolution. However, for the simulated "low-$\dot{P}$ short-P" subclass, the value of $\overline{\mathrm{FOM}}=0.45$ for the case of inclination evolution is much better than that of fixed inclination ($\overline{\mathrm{FOM}}=0.72$ ), which suggests that at least the case of inclination evolution seems to be consistent with that of observations.
  
    Moreover, the inclined evolution models of both the "high-$\dot{P}$ short-P" and "low-$\dot{P}$ short-P" subclasses have a significant increase in the initial period $\mu_{P_0}$ compared with that of the no inclination evolution model.
    The orthogonal rotator (noninclined evolution) model is not significantly dependent on $\mu_{P_0}$. In contrast, the model with inclination evolution tends to a large $\mu_{P_0}$. For the vacuum dipole model, the magnetic axis and rotation axis quickly align with each other, causing the magnetar spin-down to cease. It suggests  there is a limit on the amount of rotational energy that may be lost through magnetic dipole radiation. Therefore, the initial spin of the synthetic population cannot be significantly faster than that of the observed population, which results in the model with inclination evolution showing a $\mu_{P_0}$ dependence. What is reflected on the $P-\dot{P}$ diagram is that the evolutionary trajectory bends towards smaller $\dot{P}$ prematurely not far from the initial spin period. Therefore, the synthetic population with a short initial period is challenging to fit the observed population when inclination evolution is taken into consideration in our simulation.

    \subsection{Plasma-filled magnetosphere simulation}
\begin{figure*}
    \includegraphics[width=7in]{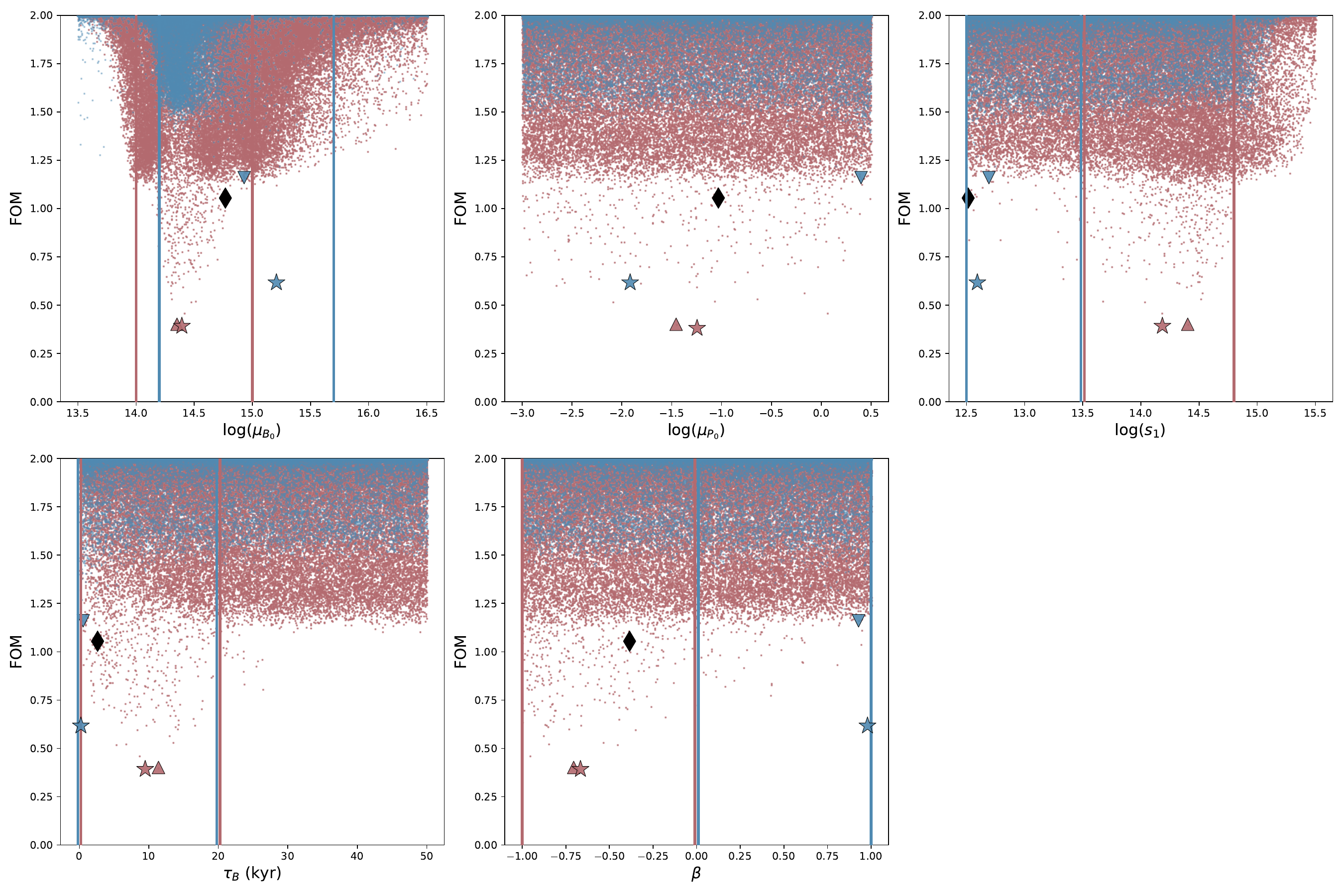}	
    \caption{Similar to the Fig.~\ref{Fig4}, the values of $\overline{\mathrm{FOM}}$ for different free parameters via first cycle simulation without inclination evolution but in plasma-filled magnetosphere.}
    \label{Fig7}
\end{figure*}

	In this section, we investigate the spin evolution of magnetar populations in a plasma-filled magnetosphere instead of vacuum magnetosphere by considering both inclination evolution and no inclination evolution. We also focus on simulating the respective distribution of the "high-$\dot{P}$ short-P" and "low-$\dot{P}$ short-P" magnetar subclasses undergoing evolution in the $P-\dot{P}$ diagram. By adopting a similar method to Sec. 5.1, we first run the automatic algorithm for the model without inclination evolution. The values of $\overline{\mathrm{FOM}}$ for different free parameters of $\mu_{B_0}$, $\mu_{P_0}$, $s_1$, $\beta$, and $\tau_B$ in the first cycle are shown in Fig.~\ref{Fig7}. We find that the parameter-dependent nature of the fixed-inclination model for the two subclasses of magnetars evolving within the plasma-filled magnetosphere is the same as that of the model evolving in a vacuum. Hence, we deduce the range of other parameters and run the next cycle until the low-$\overline{\mathrm{FOM}}$ region is no longer seen. The red and blue stars in Fig.~\ref{Fig7} correspond to the best-fit parameters with minimum $\overline{\mathrm{FOM}}$ for the "high-$\dot{P}$ short-P" and "low-$\dot{P}$ short-P" subclasses in the final cycles.
    
\begin{figure*}
    \includegraphics[width=7in]{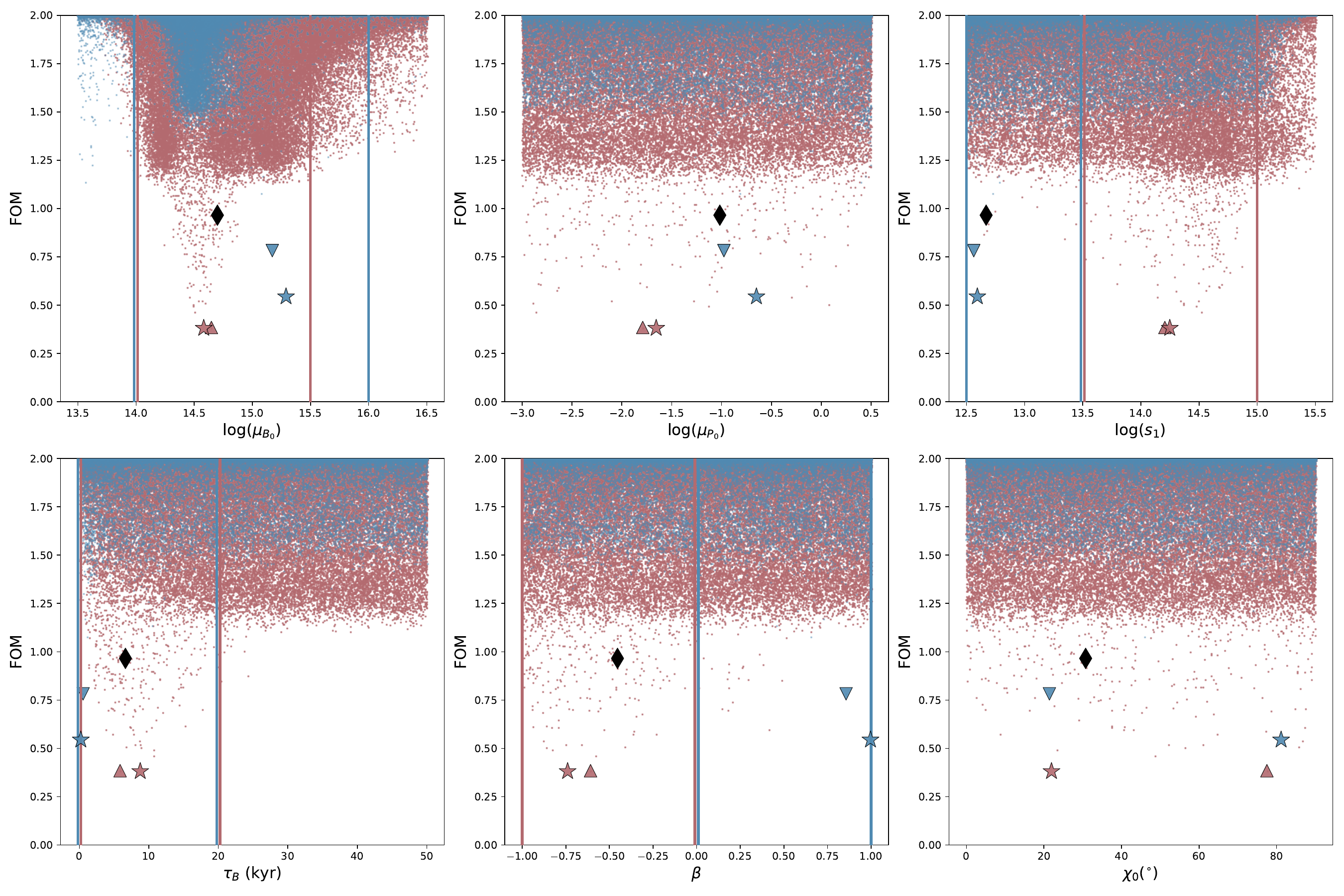}
    \caption{Similar to Fig.~\ref{Fig7}, the values of $\overline{\mathrm{FOM}}$ for different free parameters via first cycle simulation in the plasma-filled magnetosphere but considering the evolution of inclination.}
    \label{Fig8}
\end{figure*}

    Within the scenario of inclination evolution, one more free parameter (i.e., $\chi_0$) is added in the simulations to compare to the models without inclination evolution. Figure \ref{Fig8} shows the results of the first cycle, and reveals the values of $\overline{\mathrm{FOM}}$ for the different free parameters $\mu_{B_0}$, $\mu_{P_0}$, $s_1$, $\tau_B$, and $\chi_0$ in the final cycles. It is found that the "high-$\dot{P}$ short-P" and "low-$\dot{P}$ short-P" subclasses are insignificantly dependent on both $\mu_{P_0}$ and $\chi_0$, and the variations in these two parameters do not obviously affect the goodness of fit. If this is the case, we will limit the range of $\mu_{B_0}$, $s_1$, $\tau_B$ and $\beta$ parameters in the subsequent cycles. The stars marked in Fig.~\ref{Fig8} show the best-fit parameters that correspond to the minimum $\overline{\mathrm{FOM}}$ in the final cycles.
    
\begin{figure*}
    \includegraphics[width=3.5in]{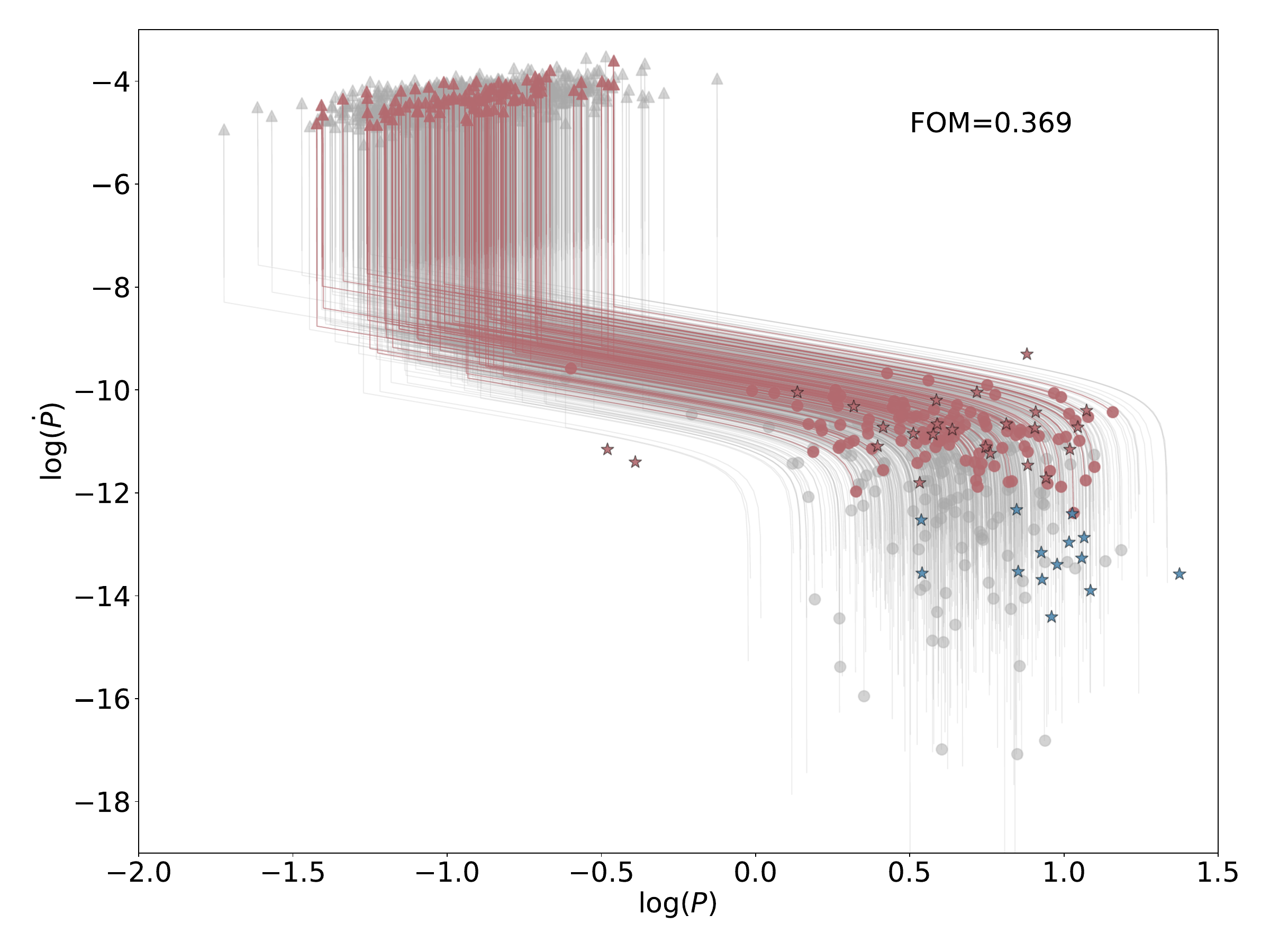}
    \includegraphics[width=3.5in]{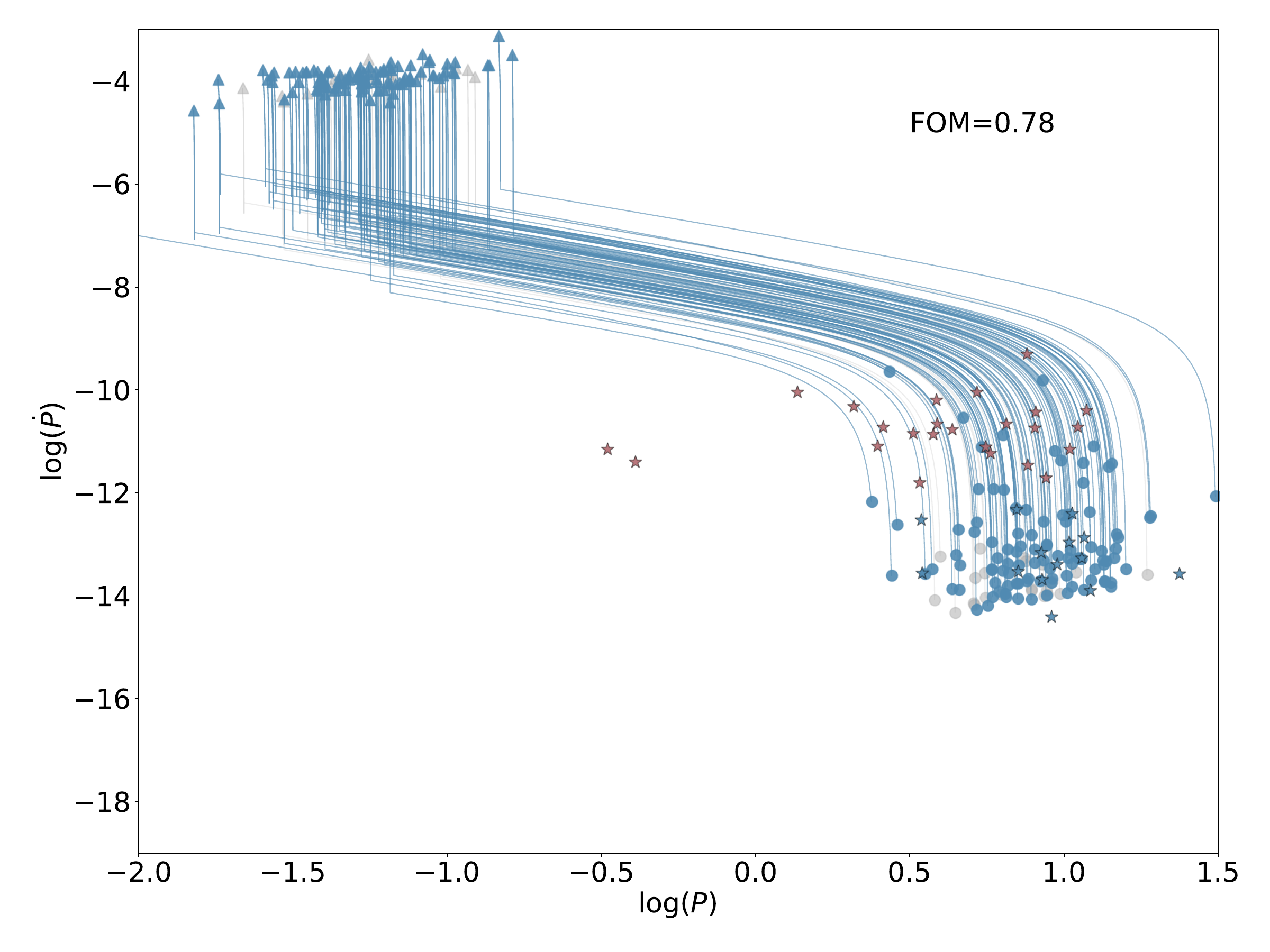}\\
    \includegraphics[width=3.5in]{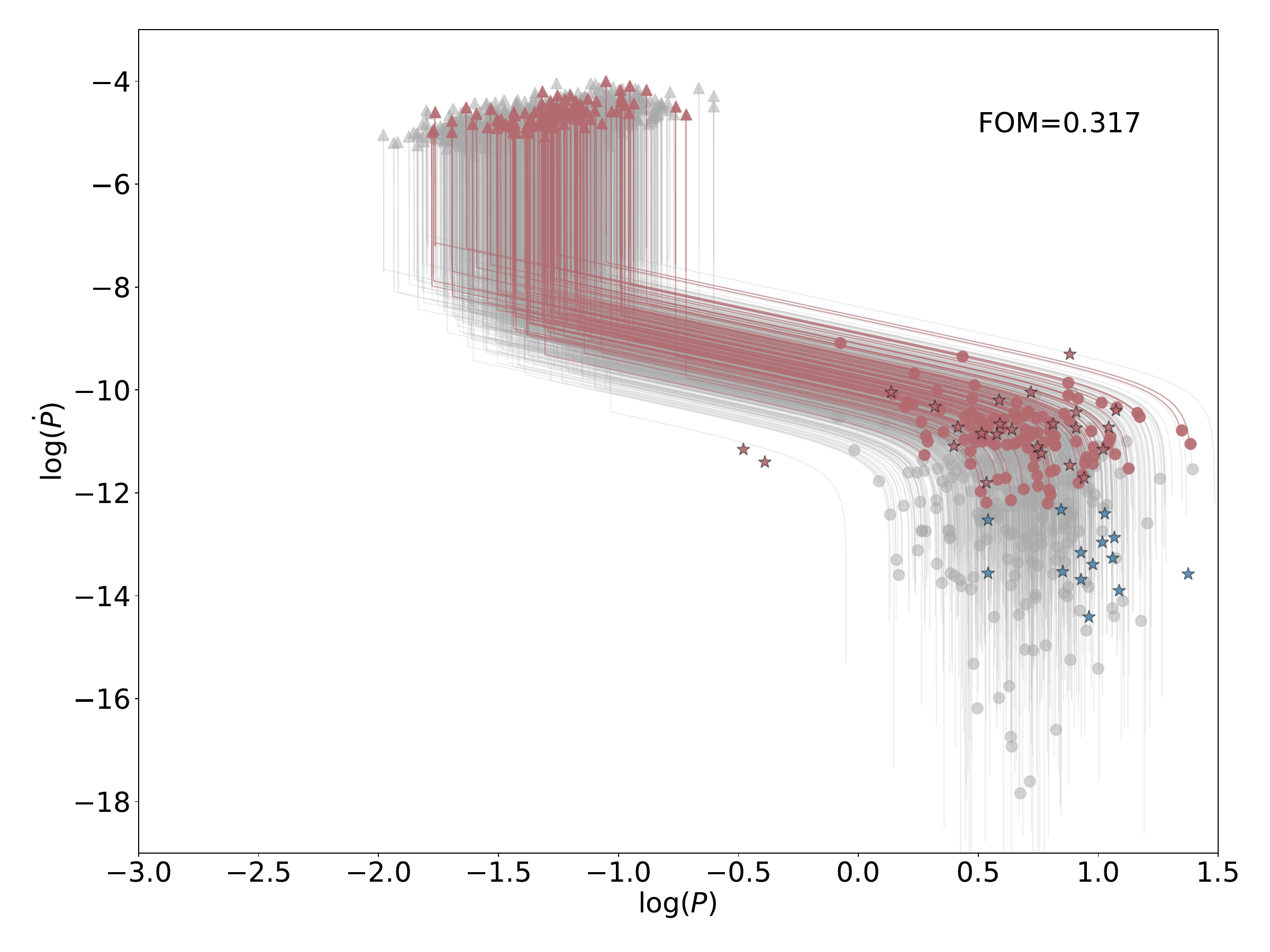}
    \includegraphics[width=3.5in]{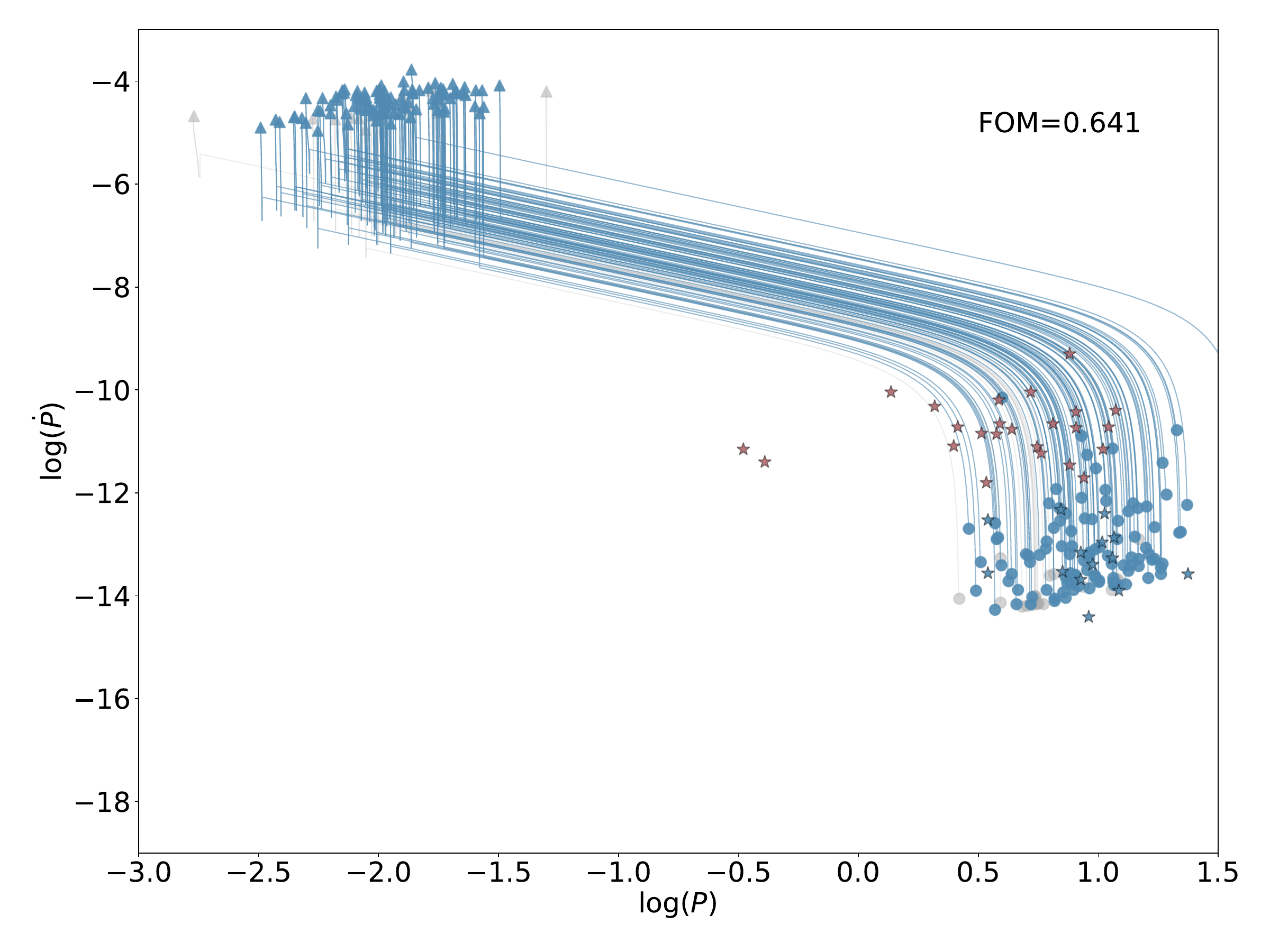}
    \caption{Similar to Fig.~\ref{Fig6}, but in plasma-filled magnetosphere.}
    \label{Fig9}
\end{figure*}

    By adopting the above best-fit parameters reported in Table \ref{tab:parameter}, we simulate the evolution tracks from a synthetic magnetar to observational magnetar populations (see Fig.~\ref{Fig9}). Similar to Fig.~\ref{Fig6}, but in a plasma-filled magnetosphere, it is found that the simulated magnetar population is mixed with that of observations excepting the fade-away population. Also, the minimum $\overline{\mathrm{FOM}}$ of "high-$\dot{P}$ short-P" subclass is as low as about 0.5, which means that the simulation result seems to be consistent with the observed magnetar population whether or not we consider the inclination evolution. However, for the "low-$\dot{P}$ short-P" subclass, the value of $\overline{\mathrm{FOM}}$ with the case of inclination evolution is lower than that without inclination evolution. (see Table.~\ref{tab:parameter}). In addition, in comparison to the case in vacuum, the initial evolution of $\dot{P}$ is rapid decline which is caused by the energy dissipation rate of neutrinos prior to 40 s.

	\section{Conclusion and Discussion}
    Magnetars are believed to be energy sources or central engines of many transients in the Universe, such as gravitational waves, super luminous supernovae, GRBs, and FRBs \citep{Kaspi.et.al.2017}. In general, the period of newborn magnetars is about several milliseconds (called millisecond magnetars), while the period of observed magnetars is longer than 2 sec \citep{Olausen.et.al.2014}. One basic question is how will magnetars evolve after they are born? In this paper, by adopting Monte Carlo simulations to generate synthetic magnetar populations, we investigate the spin and inclination evolution, as well as the decay of the magnetic field of magnetar populations in both vacuum and plasma-filled magnetospheres, and try to determine the reasons for magnetar evolution. We find the following interesting results.
    
    \begin{itemize}	
         \item[$\bullet$] By performing a statistical analysis of $P$ and $\dot{P}$ for all magnetars or candidates, it is found that the periods of all magnetars or candidates appear as a bimodal distribution with a separation line at $P\sim$68 sec. We define the two regions as “long-P” and “short-P” magnetar subclasses, respectively. Moreover, for the “short-P” magnetar subclass, we find that $\dot{P}$ also has a bimodal distribution with dividing line at $\dot{P}\sim 7.5\times 10^{-13}$, which we name the “high-$\dot{P}$ short-P” and “low-$\dot{P}$ short-P” magnetar subclasses.
        \item[$\bullet$] For the "short-P" magnetar subclass, we investigate all evolutionary models to fit the observed data. No optimal set of parameters could be found to align with observations. It suggests that the short-P magnetar subclass seems not to be a homogeneous class sharing identical evolution channels or initial properties.
        \item[$\bullet$] For the “high-$\dot{P}$ short-P” magnetar subclass, based on the value of $\overline{\mathrm{FOM}}$, we find that the simulated magnetar population is mixed with that of observations excepting the fade-away population by considering the evolution of spin and magnetic field. This means that the magnetar evolution is dependent on both spin and magnetic field, but seems to be not dependent on inclination evolution and the magnetospheric environment (vacuum or plasma-filled).
        \item[$\bullet$] For the “low-$\dot{P}$ short-P” magnetar subclass, we find that the simulated magnetar population is also mixed with that of observations excepting the fade-away population by considering the evolution of spin and magnetic field. In comparison with the case of “high-$\dot{P}$ short-P”, the magnetar evolution is dependent on the spin, magnetic field, inclination evolution, as well as the magnetospheric environment. Based on the value of $\overline{\mathrm{FOM}}$, the best evolution model should be the case of inclination evolution in vacuum.
    \end{itemize}
     
      By comparing the fitting parameters between “high-$\dot{P}$ short-P” and “low-$\dot{P}$ short-P” subclasses (in Table 2) with inclination evolution or not, it is noticed that the value of $\mu_{B_0}$ for the “high-$\dot{P}$ short-P” subclass is much less than that of the “low-$\dot{P}$ short-P” subclass, while its corresponding value of $\tau_B$ is much larger than that of the “low-$\dot{P}$ short-P” magnetar subclass. Moreover, the $\beta$ for the “high-$\dot{P}$ short-P” subclass is negative, while for the “low-$\dot{P}$ short-P” subclass is close to positive 1. The differences in those fitting results also suggests that the short-P magnetar subclass does not appear to be a homogeneous class.

     Comparing the case in vacuum with that of plasma-filled magnetospheres, the initial evolution of $\dot{P}$ is a rapid decline in the plasma-filled magnetosphere. This may be caused by the energy dissipation rate of neutrinos at the beginning of the newborn {magnetar's life}. Moreover, for both the “high-$\dot{P}$ short-P” and “low-$\dot{P}$ short-P” magnetar subclasses, the best-fit value of $\mu_{B_0}$ in the vacuum is a little bit larger than that of the plasma-filled magnetosphere.

	In the presence of inclination evolution, the hypothesis in vacuum requires a longer initial period, and brings the synthetic population to be close to the observed one. For a newborn magnetar with an initial period of a millisecond, it would be challenging to brake it to the (2-12) s period of observation population before the magnetic and rotational axes are perfectly aligned (i.e., $\dot{P}=0$). However, based on the original model in \cite{Duncan.et.al.1992} and \cite{Thompson.et.al.1993}, a short initial period is necessary for efficient mixing in early magnetars, and amplifying the magnetic field through vigorous dynamo action. This difference suggests that other braking mechanisms may play an important role and brake the early magnetars to a period of approximately 0.5 s, such as a fall-back disk that surpasses the influence of magnetic dipole radiation. 

   On the other hand, the braking index ($n$) of a magnetar appears to differ from the dipole radiation dominated with $n = 3$ of a neutron star. By considering magnetars that are potentially associated with supernova remnants and their ages, one can estimate the magnetar braking index. Some magnetars may exhibit higher braking indices with $n>3$ (e.g., 1E 2259+586, 1E 1841-045 and SGR 0501+4516) which is caused by the dipolar field decay or the external magnetospheric braking, while others may be present lower braking indices with $1<n<3$ (e.g., SGR 0526-66, SGR 1627-41, PSR J1622-4950 and CXOU J171405.7-381031) which is caused by wind-aided braking \citep{Gao.et.al.2016}.

    For the plasma-filled magnetosphere model, we assume that a stable plasma-filled environment occurs throughout its evolution in our simulations. In fact, if the surface magnetic field of neutron stars is decreasing with time, the plasma would eventually fall back onto the surface of the star when the magnetic pressure of plasma cannot support gravitational collapse. If this is the case, it can gradually result in the spin evolution switching from a plasma-filled form to a vacuum form. However, the absence of a comprehensive model for the evolution of plasma-filled magnetosphere, along with a limited number of observed objects and a multitude of parameters, makes it difficult to accurately estimate its effect on best-fit parameters. So we do not consider this situation in our simulations. In the future, possible detection of newborn magnetars by gravitational-wave observatories has the potential to impose new constraints on the properties of these isolated objects \citep{Radice.et.al.2019}. Moreover, a small fraction of magnetars in low-$\dot{P}$” subclass are only candidates, and it remains the possibility that the candidates are not magnetars at all. Our conclusions strongly depend on the classification of magnetars based on the $\dot{P}$. If those candidates are indeed not magnetar, our conclusions maybe need to consider again. 
 
    \begin{acknowledgments}
        This work is supported by the Guangxi Science Foundation the National (Grant No. 2023GXNSFDA026007), the Natural Science Foundation of China (Grant No. 11922301 and No. 12133003), and the Program of Bagui Scholars Program (LHJ).
    \end{acknowledgments}

\begin{table*}
    \centering
    \caption{Spin, magnetic field, and characteristic page of all observed magnetars are taken from \protect\cite{Olausen.et.al.2014}\protect\footnotemark. The magnetar candidates marked as "${\dagger}$" are collected from \cite{Beniamini.et.al.2023, Kulkarni.Kerkwijk.1998, Kerkwijk.Kulkarni.2001, Speagle.et.al.2011, Tan.et.al.2018, Pires.et.al.2019, Morello.et.al.2020, Rigoselli.et.al.2021}. The magnetic field and characteristic age are inferred as $B_\mathrm{obs} = 3.2\times10^{19}\sqrt{P\dot{P}}$ G and $P/2\dot{P}$, respectively.}
    \label{tab:observation}
    \begin{tabular}{lcccc}
        \hline \hline
        \textbf { Magnetars } & \makecell[c]{$P$\\$(\mathrm{s})$} & \makecell[c]{$\log (\dot{P})$\\$\left(\mathrm{s}~\mathrm{s}^{-1}\right)$} & \makecell[c]{$\log \left(B_{\mathrm{obs}}\right)$\\$(\mathrm{G})$} & \makecell[c]{$ \mathrm{Age} $\\$ (\mathrm{kyr}) $}\\ 
        \hline \hline
        \text { PSR J1846-0258 } & 0.33 & -11.15 & 13.69 & 0.73\\
        \text { PSR J1119-6127 } & 0.41 & -11.40 & 13.61 & 1.61\\
        \text { Swift J1818.0-1607 } & 1.36 & -10.04 & 14.55 & 0.24 \\
        \text { 1E 1547.0-5408 } & 2.07 & -10.32 & 14.50 & 0.69\\
        \text { Swift J1834.9-40846 } & 2.48 & -11.10 & 14.15 & 4.9 \\
        \text { SGR 1627-41 } & 2.60 & -10.72 & 14.35 & 2.2\\
        \text { SGR 1935+2154 } & 3.25 & -10.84 & 14.32 & 3.6\\
        \text {RX J1605+3249$^{\dagger}$ } & 3.39 & -11.80 & 13.87 & 33.59 \\
        \text {PSR J0726-2612$^{\dagger}$ } & 3.44 & -12.53 & 13.51 & 181.8 \\
        \text {RX J0420-5022$^{\dagger}$ } & 3.45 & -13.56 & 13.00 & 1983 \\
        \text { SGR J1745-2900 } & 3.76 & -10.86 & 14.36 & 4.3\\
        \text { CXOU J171405.7-381031 } & 3.83 & -10.19 & 14.70 & 0.95\\
        \text { Swift J1555.2-5402 } & 3.86 & -10.66 & 14.46 & 2.78\\
        \text { PSR J1622-4950 } & 4.33 & -10.77 & 14.44 & 4.0\\
        \text { SGR 1900+14 } & 5.20 & -10.04 & 14.85 & 0.9\\
        \text { XTE J1810-197 } & 5.54 & -11.11 & 14.32 & 11\\
        \text { SGR 0501+4516 } & 5.76 & -11.23 & 14.27 & 15\\
        \text { 1E 1048.1-5937 } & 6.46 & -10.65 & 14.59 & 4.5\\
        \text { 1E 2259+586 } & 6.98 & -12.32 & 13.77 & 230\\
        \text {RX J1856-3754$^{\dagger}$ } & 7.06 & -13.53 & 13.17 & 3728  \\
        \text { SGR 1806-20 } & 7.55 & -9.31 & 15.29 & 0.24\\
        \text { SGR 1833-0832 } & 7.57 & -11.46 & 14.22 & 34\\
        \text { CXOU J010043.1-721134 } & 8.02 & -10.73 & 14.59 & 6.8\\
        \text { SGR 0526-66 } & 8.05 & -10.42 & 14.75 & 3.4\\
        \text {RX J0720-3125$^{\dagger}$ } & 8.39 & -13.15 & 13.39 & 1905 \\
        \text { Swift J1822.3-1606 } & 8.44 & -13.68 & 13.13 & 6300\\
        \text { 4U 0142+61 } & 8.69 & -11.69 & 14.13 & 68\\
        \text { SGR 0418+5729 } & 9.08 & -14.40 & 12.79 & 36000\\
        \text {RX J2143+0654$^{\dagger}$ } & 9.43 & -13.39 & 13.30 & 3646  \\
        \text {RX J1308+2127$^{\dagger}$ } & 10.31 & -12.95 & 13.54 & 1486 \\
        \text { Swift J1830.9-0645 } & 10.42 & -11.15 & 14.74 & 23.34\\
        \text { CXOU J164710.2-455216 } & 10.61 & -12.40 & 13.82 & 420\\
        \text { 1RXS J170849.0-400910 } & 11.01 & -10.71 & 14.67 & 9.0\\
        \text {RX J0806-4123$^{\dagger}$ } & 11.37 & -13.25 & 13.40 & 3278 \\
        \text { 3XMM J185246.6+003317 } & 11.56 & -12.85 & 13.61 & 1300\\
        \text { 1E 1841-045 } & 11.79 & -10.39 & 14.85 & 4.6\\
        \text {PSR J2251-3711$^{\dagger}$ } & 12.12 & -13.89 & 13.10 & 14782 \\
        \text {PSR J0250+5854$^{\dagger}$} & 23.54 & -13.57 & 13.41 & 13739  \\
        \hline
        \text {PSR J0901-4046$^{\dagger}$} & 76 & -12.66  & 14.41 & 5347  \\
        \text {SGR 0755-2933$^{\dagger}$} & 308 & - & - & - \\
        \text {SXP 1062$^{\dagger}$} & 1070 & -5.52 &  18.26  &  0.0057  \\
        \text {GLEAM-X J1627$^{\dagger}$}  & 1091 & $<-9$ & $\lesssim 16.52$ & $\gtrsim 17.30$ \\
        \text {GCRT J1745-3009$^{\dagger}$ } & 4630 & - & - &  -  \\
        \text {4U 2206+54$^{\dagger}$} & 5750 & -6.22 & 18.27  & 0.15 \\
        \text {IGR J16358-4726$^{\dagger}$} & 5970 & - & - & - \\
        \text {4U 0114+65$^{\dagger}$} & 9350 & - & - & - \\
        \text {4U 1954+319$^{\dagger}$} & 20500 & - & - & - \\
        \text {1E  161348-5055$^{\dagger}$} & 24030.42 & <-8.80 & <17 & >544 \\
        \text {AX J1910.7+0917$^{\dagger}$ } & 36000 & - & - & - \\
        \text {FRB 20180916B (R3)$^{\dagger}$} & 1.4 $\times 10^{6}$ & - & - & - \\
        \hline \hline
    \end{tabular}
    \begin{center}
        \footnotetext{\protect\url{http://www.physics.mcgill.ca/~pulsar/magnetar/main.html}.}
    \end{center}
\end{table*}

\begin{table*}
    \centering
    \caption{
        Best-fit results of parameters for automatic algorithm. The values of "short-P" subclass are the best-fit results of the first cycle, corresponding to the values marked with black diamonds in Figs~.\ref{Fig4}, \ref{Fig5},\ref{Fig7}, and \ref{Fig8}. And the values of "high-$\dot{P}$ short-P" and low-$\dot{P}$ subclasses are the best-fit results of their respective final cycle, corresponding to the values marked with red and blue stars in these figures, respectively.}
    \label{tab:parameter}
    \begin{threeparttable}
        \begin{tabular}{ccccccccccc}
            \hline \hline
            \textbf{Classes} &\multicolumn{2}{c}{\textbf{Evolutionary  Model}} & \makecell[c]{$\mu_{B_0}$\\$(10^{14}~\mathrm{G})$} &  \makecell[c]{$\mu_{P_0}$ \\ $(\mathrm{s})$}& \makecell[c]{$s_1$\\$(10^{14}~\mathrm{G})$} & \makecell[c]{$\tau_B$\\$(\mathrm{kyr})$} & $\beta$ & \makecell[c]{$\chi_0$\\$(^{\circ})$} &$\overline{\mathrm{FOM}}$\tnote{$*$}& \makecell[c]{$\overline{\mathrm{BR}}$\\$(\mathrm{kyr}^{-1})$}\\
            \hline \hline
            \multirow{4}{*}{\textbf{short-P}} &\multirow{2}{*}{\textbf{Vacuum}}& \text{fix inclination}& 
             9.69 & 0.0171  & 0.0530  &  3.12 & -0.30  & 90.00 & 1.10$\pm$0.28 &4.75$\pm$0.43 \\&
            &\text{inclination evolution}& 
             7.26 & 2.66  & 13.10 & 9.64  & -0.02  & 74.04  & 1.22$\pm$0.29 &22.32$\pm$3.80 \\
            \cmidrule{2-11}
            &\multirow{2}{*}{\textbf{Plasma-filled}}&\text{fix inclination}& 
            5.86 &  0.0929 & 0.0327 & 2.63 & -0.38 & 90.00 & 1.13$\pm$0.28 & 6.60$\pm$0.63 \\&
            & \text{inclination evolution}& 
            5.01 & 0.0960 & 0.0467 & 6.63 &-0.45 & 30.74 &1.15 $\pm$0.30 &3.05$\pm$ 0.26\\
            \hline \hline
            \multirow{4}{*}{\textbf{high-$\dot{P}$ short-P}} &\multirow{2}{*}{\textbf{Vacuum}}& \text{fix inclination}& 
            4.40 & 0.0302\tnote{$\dagger$} & 2.30 & 9.12 & -0.81 & 90.00 & $0.50\pm0.28$ & $6.03\pm0.67$ \\&
            &\text{inclination evolution}& 
            4.86 & 0.375 &1.67 & 8.72 & -0.90 & 89.81 & $0.49\pm0.27$ & $5.08\pm0.52$ \\
            \cmidrule{2-11}
            &\multirow{2}{*}{\textbf{Plasma-filled}}&\text{fix inclination}& 
            2.48& 0.0568\tnote{$\dagger$}&1.53& 9.48&-0.67&90.00 &0.52$\pm$ 0.29& 3.61$\pm$0.40 \\&
            & \text{inclination evolution}&3.82 & 0.0221\tnote{$\dagger$} &
            1.77 & 8.76 & -0.74 & 21.93\tnote{$\dagger$} & $0.52\pm0.28$ & $5.55\pm0.56$ \\
            \hline 
            \multirow{4}{*}{\textbf{low-$\dot{P}$ short-P}} &\multirow{2}{*}{\textbf{Vacuum}}&\text{fix inclination}& 
            28.65 & 0.00180\tnote{$\dagger$}  &0.0448 & 0.207 & 0.99& 90.00& $0.72\pm0.25$ & $0.31\pm0.01$ \\&
            &\text{inclination evolution}& 
            24.44 & 3.45  & 0.0510 &0.403  & 1.00 & 79.01 & $0.45\pm0.22$ & $0.30\pm0.01$ \\
            \cmidrule{2-11}
            &\multirow{2}{*}{\textbf{Plasma-filled}}& \text{fix inclination}&
             16.10&  0.0121\tnote{$\dagger$}&0.0335 &0.22 & 0.98& 90.00& 0.73$\pm$0.25 & 0.33$\pm$0.012\\&
            &\text{inclination evolution}& 
            19.47 & 0.224\tnote{$\dagger$} & 0.0392 & 0.209 & 1.00 & 81.16 & 0.63$\pm$0.25 & 0.32 $\pm$0.01 \\
            \hline \hline
        \end{tabular}
        \begin{tablenotes}
            \item [*] Note: The $\overline{\mathrm{FOM}}$ presented here is the result of an average of $\mathrm{N_{pop}}=1000$ calculations, differing from the values marked as the stars in Fig.~\ref{Fig4}, \ref{Fig5}, \ref{Fig7}, and \ref{Fig8}, which show the average results from only $\mathrm{N_{pop}}=50$ calculations.
            \item [$\dagger$] These parameters do not exhibit any discernible influence on $\overline{\mathrm{FOM}}$ across all the cycles.
        \end{tablenotes}
    \end{threeparttable}
\end{table*}


\bibliography{ms}

\end{document}